\documentstyle[aps,eqsecnum]{revtex}

\newcommand{\nub}{{\overline{\nu}}}
\newcommand{\dM}[1]{\partial_{\mu_{#1}}}

\newcommand{\dE}[1]{\partial_{\eta_{#1}}}
\newcommand{\dNb}[1]{\partial_{\overline{\nu}_{#1}}}
\newcommand{\dmi}{\partial_{\mu_i}}

\newcommand{\dmj}{\partial_{\mu_j}}

\newcommand{\dej}{\partial_{\eta_j}}

\newcommand{\tr}{{\rm tr}}
\newcommand{\Sn}{S(\nu,\mu)}
\newcommand{\Q}[1]{Q_{{#1}\times {#1}}}

\newcommand{\Gb}{\Psi}
\newcommand{\dm}{\partial_\mu}
\newcommand{\dn}{\partial_\nu}
\newcommand{\de}{\partial_\eta}
\newcommand{\bx}{{\bf x}}
\newcommand{\bA}{{\bf A}}
\newcommand{\bb}{{\bf b}}
\newcommand{\ri}{{\rm i}}
\newcommand{\re}{{\rm e}}
\newcommand{\bQ}{{\bf Q}}
\newcommand{\bR}{{\bf R}}
\newcommand{\dnmS}[1]{\partial_{\nub_{#1}}\partial_{\mu_{#1}}S_{#1}}
\newcommand{\dmmS}[1]{\partial_{\mu_{#1}}^2S_{#1}}
\newcommand{\dnnS}[1]{\partial_{\nub_{#1}}^2S_{#1}}
\newcommand{\mL}{\overline{M}^{(l)}}
\newcommand{\ml}{\overline{M}_l}
\begin{document}
\draft

\title{Semiclassics for a Dissipative Quantum Map}
\author{Daniel Braun$^{(1)}$, Petr A.~Braun$^{(1,2)}$, Fritz Haake$^{(1)}$}
\address{$^{(1)}$ FB7, Universit\"at--GHS Essen, 45\,117 Essen, Germany\\
$^{(2)}$ Department of Theoretical Physics, Institute of Physics,
Saint-Petersburg University, Saint-Petersburg 198904, Russia}

\maketitle
\thispagestyle{empty}
We present a semiclassical analysis for a dissipative quantum map with an
area--nonpreserving classical limit. We show that in the limit  $\hbar\to
0$ the trace of an arbitrary natural power of  the  
propagator is dominated by contributions from  periodic orbits of the
corresponding classical dissipative motion. We  derive 
trace formulae of the Gutzwiller type for such quantum maps. In comparison
to 
Tabor's formula  for area--preserving maps, both classical action and
stability 
prefactor are modified by the dissipation.  We evaluate the traces
explicitly in the case of a dissipative kicked top  with integrable
classical motion and find good agreement with numerical results.
\bigskip

\section{Introduction}\label{intro}
A good deal of what we know about the quantum mechanics of classically
chaotic systems is due to Gutzwiller's trace formula \cite{Gutz70,Gutz71}. It
expresses the fluctuation of the density of states to lowest order in
Planck's constant $\hbar$ by a sum over classical periodic orbits.  Gutzwiller
developed this semiclassical formalism for conservative Hamiltonian
systems. Later 
on, the formalism was applied by Tabor to
area--preserving maps \cite{Tabor83}.
A lot is known about the quantum mechanical as well as the
classical dynamics of dissipative chaotic systems. While the quantum energy
spectra of conservative 
chaotic systems tend to have spectral statistics like the well known
ensembles of random Hermitian matrices \cite{Bohigas84}, the spectrum of the
quantum mechanical propagator of dissipative chaotic systems seems to obey
the statistics  of Ginibre's ensemble. This was shown
numerically by analyzing  a damped version of the kicked top
\cite{Grobe88}. The 
eigenphases become complex and the spacing distribution $P(s)$ of nearest
eigenvalues  shows a universal small--$s$ behavior, $P(s)\propto s^3$, as soon
as the  
dissipation is strong enough. For a classically integrable
dissipative system, $P(s)\propto s$. This is at variance with the result for
conservative systems, with the $P(s)$ of Wigner and Dyson, $P(s)\propto
s^\beta$ ($s\ll 1$) with $\beta=1,2,4$ depending on the 
time-- and spin--reversal symmetry of the system. For classically integrable
conservative 
systems, the spectrum is uncorrelated and $P(s)=\exp(-s)$
\cite{Rosenzweig60}. \\ 
Classical dissipative and chaotic
dynamics typically leads to phase--space structures dominated by strange
attractors, i.e. attractors of fractal (or multifractal) dimension,
whereas for conservative systems initial distributions in phase 
space rapidly converge towards a stationary uniform one on the energy
shell\cite{Ott93}. 
 
Particularly noteworthy earlier semiclassical work on dissipative processes
is the quantization of Henon's map with dissipation  by
Graham and T\'el
\cite{Graham85}; Dittrich and Graham considered the kicked rotator
with dissipation \cite{Dittrich85}.   In \cite{Eckhardt94} a
periodic--orbit analysis of the classical spectral
density for the Lorenz model, the Hausdorff dimension of the
Lorenz attractor, and the average Lyapunov exponent was put
forward. Refs.\cite{Peplowski91} 
and \cite{Iwaniszewski95}  treat  numerically the  
transition from quantum mechanics to classical mechanics
 for the kicked top with dissipation. In \cite{Peplowski91} it was shown by
direct numerical diagonalization of  the propagator of the density matrix
that the eigenstate corresponding to the lowest eigenvalue when written in a
coherent state basis tends towards the corresponding stationary classical
distribution function, in particular to (a smoothed out version) of the
strange attractor in the case of classically chaotic motion. In
\cite{Iwaniszewski95} the same analysis was carried further by solving the
equivalent stochastic Schr\"odinger equation. This procedure allowed for
much larger values of the angular momentum and correspondingly smaller
values of $\hbar$ and showed again a remarkable agreement between the two
stationary distributions in phase space for the quantum and classical
case. \\

In this paper we establish a periodic-orbit theory for a dissipative
quantum map $P$ that consists of a purely unitary step followed by a purely
dissipative one, the dissipative one describable by a Markovian master
equation.  We show that to leading order in $\hbar$ traces of
arbitrary natural powers of $P$ are given by  
a Gutzwiller type formula,
\begin{equation} \label{trPNintro}
\tr P^N=\sum_{p.p.}\frac{\re^{J\sum_{i=1}^N R_i}}{\left|\tr
\prod_{i=N}^1M_d^{(i)}-\tr 
M\right|}\mbox{, }\quad\quad N=1,2,\ldots\,,
\end{equation}
where the sum is over all periodic points of the corresponding classical map
$P_{cl}^N$ in phase space. The action appearing in the exponent is a classical
action from the dissipative steps in $P_{cl}^N$, the matrices $M_d^{(i)}$
are the monodromy matrices for the dissipative steps and $M$ is the total
monodromy matrix of $P_{cl}^N$. The inverted order of the indices at the
product indicates that the matrices 
are ordered from left to right according to decreasing indices. We will
discuss (\ref{trPNintro}) in detail 
below and illustrate our
reasoning for a dissipative and periodically driven top, which in the
integrable case allows for explicit evaluation of all traces.  

By tops \cite{kickedtop,Haake91} we mean systems whose dynamical variables
are the three components of an angular momentum, $J_{x,y,z}$, the squares of
which have a conserved sum, ${\bf J}^2=j(j+1)$. In the classical limit
(formally attained by letting the quantum number $j$ approach infinity) the
surface of the unit sphere $\lim_{j\rightarrow\infty} ({\bf J}/j)^2=1$
becomes the phase space, such that one confronts but a single degree of
freedom. Classical nonintegrability can if desired be enforced by periodic
driving. Quantum mechanically, one deals with a $(2j+1)$ dimensional Hilbert
space within which one must diagonalize the Floquet operator to get the
spectrum of $(2j+1)$ quasienergies. The semiclassical limit is characterized
by large values, integer or half integer, of the quantum number $j$. The
convenience of periodically driven tops for  investigations of spectral
properties mostly lies in the compactness of the classical phase space and
the corresponding finite dimensionality of the quantum mechanical Hilbert
space.

Our theory applies to physical situations where dissipation is
important. One might try experimental realisations with 
several identical two--level atoms collectively interacting with a single
mode of the electromagnetic field, as in the superradiance experiment of
Haroche \cite{Haroche}.

The paper is organized as
follows: In Section \ref{sectparts} we introduce the map to be studied, the
composition of a purely unitary and a purely dissipative part. Semiclassical
approximations for both components of the map are reviewed. 
Section \ref{secttrP} is devoted to the derivation of a trace formula for
the propagator and Section
\ref{secttrPn} deals with the generalization to higher traces. In
\ref{secKic} we apply the theory to an integrable dissipative kicked top
and  compare with numerical results, and in Section \ref{sectconcl} we
summarize our main conclusions.  

\section{The Map}\label{sectparts}

While the stroboscopic description of unitary dynamics employs a wave
function $\psi$ and a unitary Floquet operator $F$ as $\psi(nT)=F^n\psi(0)$,
a dissipative quantum map requires a density matrix  
$\rho(nT)=P^n\rho(0)$ with a nonunitary propagator $P$. For the sake of
convenience we imagine dissipative and purely unitary motion to take place
separately and in sequence during each period such that the map takes the
form  
\begin{equation} \label{eq:rhot}
\rho((n+1)T)=DF\rho(nT)F^{\dagger}\equiv P\rho(nT)\,;
\end{equation} 
in contrast to the unitary part of this map (which factorizes in $F$ and
$F^\dagger$) the dissipative part $D$ cannot be split into two factors. 
Our goal is to determine spectral properties of the total
propagator $P$, a linear, non--unitary operator, by semiclassical means. As
a preparation we briefly review what is known about semiclassical 
approximations of the two components of the map, starting with the unitary
part. 

\subsection{Semiclassics for the unitary evolution}\label{secuni}
The Floquet operator $F$ may by itself describe a motion with either an
integrable or chaotic classical limit. A simple integrable example is a pure
rotation by an angle $\beta$ about, say, the $y$-axis with $F=\exp(-{\rm
i}\beta J_y)$. Most illustrations of general results will in fact be given
for that case. If desired, a chaotic top can be treated by including a
torsion about the $z$-axis as $F=\exp(-{\rm i}\frac{k}{2j+1}J_z^2)\exp(-{\rm
i}\beta J_y)$. The corresponding classical motion proceeds in the spherical
phase space defined by the conservation law ${\bf J}^2=j(j+1)$ and may be
described by a pair of canonical variables; a convenient choice is attained
by specifying the orientation of the classical unit vector
$\lim_{j\rightarrow\infty}\frac{{\bf J}}{j}$ by a polar angle $\Theta$ with 
respect to the $z$-axis and an azimutal angle $\phi$ reckoned against the
$x$-axis in the $xy$-plane and then taking $\cos\theta$ as the momentum and
$\phi$ as the conjugate coordinate. 

A semiclassical approximation for $F$ can be derived by representing $F$ in
the $J_z$ basis ($J_z|m\rangle=m|m\rangle$). The matrix element
$F_{nm}=\langle n|F|m\rangle$ is the probability amplitude to get the
$z$-projection of the angular momentum mapped from the initial value $m$ to
the final one $n$. If $F$ describes a pure rotation that matrix element
becomes Wigner's $d$--function whose asymptotic behavior for large
$J=j+\frac{1}{2}$ is  
well known \cite{PBraun93}. The discrete quantum numbers
$m$ and $n$ can be replaced in the limit $J\to\infty$ by the  continuous
classical initial and final momenta 
\begin{equation} \label{nm}
\mu=m/J=\cos\Theta_i\mbox{, and }\nu=n/J=\cos\Theta_f\,,
\end{equation}
respectively. 
The semiclassical approximation has the familiar van Vleck form in which a
classical action $\Sn$ enters as a phase, and its second mixed derivative
\begin{equation} \label{CNM}
C(\nu,\mu)=\frac{(-1)^j}{\sqrt{2\pi J}}\sqrt{|\dn\dm 
\Sn|}
\end{equation}
as a preexponential factor. All classical orbits connecting the initial and
the final momentum contribute a term of the structure $C(\nu,\mu)\re^{{\rm
i}J\Sn}$. Explicit expressions for the function $S(\nu,\mu)$ can be found in
\cite{PBraun96}, both for integrable cases such as pure rotations and
chaotic ones involving rotations and torsions. For our present prime
example, the linear rotation by $\beta$ about the $y$-axis, there are
typically  two classical trajectories differing by the signs of  azimuth
and action;  only  
if $\beta>\theta_i+\theta_f$ or  $\beta<|\theta_i-\theta_f|$, no classical
trajectories connecting the initial with the final momentum exist;  if one
of the foregoing inequalities is turned into an equality we have
$\phi_i=\phi_f=0$ and only one trajectory contributes (see the discussion in
\cite{PBraun96}); we shall not consider these latter exceptions here but only
treat the case of two contributing trajectories for which the amplitude
under discussion reads   
\begin{equation} \label{eq:dmn}
F_{nm}\simeq\sum_{\sigma=\pm
1}C(\nu,\mu)\re^{\ri\sigma(J\Sn-\frac{\pi}{4})}\,.
\end{equation}
The action $S(\nu,\mu)$ (or $-S$ if $\phi<0$) also serves as the generating
function of the classical map    
\begin{equation}
 (\mu,\phi_i)\to (\nu,\phi_f)
\label{map}
\end{equation}
in the sense
\begin{equation} \label{eq:gen}
\dn \Sn=\mp\phi_f(\nu,\mu)\mbox{, }\dm \Sn=\pm\phi_i(\nu,\mu)\,;
\end{equation}
here, the upper (lower) sign refers to $\phi_{i,f}>0$ ($<0$); note that for the
rotation about the $y$--axis the sign of the initial and final azimuth is
the same.

The generating property (\ref{eq:gen}) and the general van Vleck form
(\ref{eq:dmn}) will allow us 
to develop semiclassical formulas largely independent of the particular
dynamics. In particular, the specific form of $\Sn$ will not be needed below.

\subsection{Semiclassics for the dissipative part of the map}\label{Scdip}

To introduce the dissipative part $D$ of our map (\ref{eq:rhot}) we
momentarily allow for a continuous time and employ the master equation of
spin damping well known from the theory of
superradiance\cite{Bonifacio71.1,Haroche,Benedict}, 
\begin{equation} \label{eq:rhotd}
\dot{\rho(t)}=\kappa([J_-,\rho(t) J_+]+[J_-\rho(t), J_+])\equiv
\Lambda\rho(t)\,,
\end{equation}
where the damping constant $\kappa$ sets the time scale on which the angular
momentum in question approaches, while keeping its square ${\bf J}^2=j(j+1)$
conserved, the stationary state of minimal $z$-component, $J_z=-j$; the
operators $J_{\pm}=J_x\pm{\rm i}J_y$ are the familiar raising and lowering
operators.  

Let us first look at the {\em classical} limit. The classical equations of
motion can be determined  by extracting equations for the
expectation values of $J_z$ and $J_\pm$, parameterizing them as $\langle
J_z\rangle =J\cos\Theta$ 
and $\langle J_\pm\rangle =J\sin\theta \re^{\pm{\rm i}\phi}$ and factorizing  all
operator products (e.g. \,$\langle J_+J_-\rangle\rightarrow\langle
J_+\rangle\langle 
J_-\rangle$).     
One then finds that the angular momentum behaves classically like
an overdamped pendulum \cite{Haake91},
\begin{equation} \label{eq:pendeom}
\dot{\phi}=0\mbox{, }\quad\quad\dot{\Theta}=2J\kappa\sin\Theta\,.
\end{equation}
The classical
damping constant is thus revealed as $2J\kappa$. For future convenience, we
give the solution of 
(\ref{eq:pendeom}) in terms of the dimensionless time $s=2J\kappa t$
and phase--space variables $\mu(s)=\cos \Theta(s)$ and $\phi$:
\begin{equation} \label{eq:sldmp}
s=\frac{1}{2}\ln\frac{(1-\mu(s))(1+\mu(0))}{(1-\mu(0))(1+\mu(s)))}\mbox{,
}\quad\quad\phi(s)=\phi(0)\,. 
\end{equation}

For the sake of definiteness we imagine the damping to be effective during the
full duration $T$ of a period, thus confining the previous unitary motion to
an instantaneous kick. The dissipative part of the map is then obtained by
exponentiating the continuous-time generator $\Lambda$ of the above master
equation as $D=\exp(\Lambda T)$. 
 
The {\em semiclassical} approximation for the dissipative propagator
$D=\exp(\Lambda T)$ looks much like the familiar van Vleck  
form for  unitary propagators. Details of the theory which
amounts to solving a Hamilton--Jacobi equation for the classical generating
function of the dissipative motion can be
found in \cite{PBraun97.2}. Here we limit
ourselves to exposing the idea and summarizing the main results. \\

Looking at the master equation (\ref{eq:rhotd}) in the $J_z$ basis we reveal
an important conservation 
law upon expressing the density matrix element $\langle
m_1|\rho(s)|m_2\rangle$ as a function of the mean $m=\frac{m_1+m_2}{2}$
and the skewness
$k=\frac{m_1-m_2}{2}$ of the quantum numbers $m_1,m_2$,
i.e. $\rho_m(k,s)=\langle 
m_1|\rho(s)|m_2\rangle$. The operators $J_\pm$ give rise to
the functions  
$d_m=j(j+1)-m(m-1)$ known from the angular momentum algebra 
($J_\pm|m\rangle=\sqrt{d_{\mp m}}|m\pm 1\rangle$) and
$f_m(k)=\sqrt{d_{m+k}d_{m-k}}$. With these abbreviations our master equation 
reads
\begin{equation} \label{eq:rhodtm}
J\frac{d\rho_m(k,s)}{ds}=f_{m+1}(k)\rho_{m+1}(k,s)-(d_m-k^2)\rho_m(k,s)\,.
\end{equation}
Obviously,  the skewness $k$
enters only as a parameter and is conserved. 
The general solution of (\ref{eq:rhodtm}) can be
written with the help of a ``dissipative propagator'' $D_{mn}(k,s)$ as 
\begin{equation} \label{eq:rhoDrho}
\rho_m(k,s)=\sum_n D_{mn}(k,s)\rho_n(k,0)\,.
\end{equation}
For large $J$, the variables $\mu=m/J$ and $\nu=n/J$ prove useful
again. A third one is introduced for $k$, $\eta=k/J$, and we write 
consequently $D_{mn}(k,s)=J^{-1}D(\mu,\nu;\eta;s)$.\\
The clue towards $D$ is a WKB type ansatz, 
\begin{equation} \label{WKB}
D(\mu,\nu;\eta;s)=B(\mu,\nu;\eta;s)\exp(JR(\mu,\nu;\eta;s))\,,
\end{equation}
and to expand systematically in $1/J$.  Note that even
though the functions
$B$ and $R$ are assumed smooth, the WKB form still allows for a delta peak
as $J\to \infty$.  To lowest order
in $1/J$, (\ref{eq:rhoDrho}) leads to an equation for $R$ which looks
like a Hamilton--Jacobi equation for a one dimensional Hamiltonian system
with the somewhat 
unconventional Hamilton function $H(\mu,p)=(1-\mu^2)\left(1-\exp(p)\right)$,
\begin{equation} \label{HJ}
\frac{\partial R}{\partial s}+H(\mu ,\frac{\partial R}{\partial \mu})=0\,.
\end{equation}
In this Hamiltonian description of the semiclassical approximation of our
damped spin the variable $\mu$ appears as a coordinate with a canonically
associated momentum $p=\frac{\partial R}{\partial \mu}$. The Hamilto--Jacobi
equation (\ref{HJ}) is equivalent to two Hamilton equations  
of motion for $\mu$ and $p$. The solutions involve two integration 
constants, one of which we 
choose as the initial coordinate $\nu$; as the second will serve the
conserved ``energy'' $E=H(\mu,p)$ which for the sake of convenience we
represent by the parameter $a=\sqrt{1-E}$. With 
this choice, we get the Hamiltonian trajectories
\begin{equation} \label{eq:sped}
s=\frac{1}{2a}\ln \frac{(\nu+a)(\mu-a)}{(\nu-a)(\mu+a)}\,.
\end{equation}
Comparison with the classical dynamics (\ref{eq:sldmp}) shows that $\mu$ and
$\nu$ are connected 
by a classical trajectory iff $a=1$. If one thinks of a Hamiltonian
trajectory as specified by the initial and final coordinates $\nu$ and $\mu$
with a given time span $s$, as we shall frequently do below, one should read
(\ref{eq:sped}) as determining the parameter $a$ as a function 
$a=a(\mu,\nu,s)$. The 
action $R$ can  be obtained by integrating $pd\mu$ along the 
trajectory (\ref{eq:sped}) and reads \cite{PBraun97.2}
\begin{eqnarray} \label{eq:R}
R(\mu,\nu;\eta;s)&=&\frac{1}{2}(\xi(1,\nu-\eta)-\xi(1,\mu-\eta)+\xi(1,\nu+\eta)-\xi(1,\mu+\eta))\nonumber\\
&&-\xi(a,\nu)+\xi(a,\mu)+s(a^2-1+\eta^2)\,,\nonumber\\
\xi(x,y)&\equiv&(x+y)\ln(x+y)-(x-y)\ln(x-y)\,.
\end{eqnarray}
Several features of the action $R$ will be of importance in the
sequel:
\begin{itemize}
\item For $\eta=0$ we get $a=1$ from $\dm R=0$, thus
the classical equation of motion (\ref{eq:sldmp}); then $\dn R=0$
holds as well.
\item For $\eta=0$  and $\mu$, $\nu$ connected by the
classical trajectory (i.e. $a=1$), $R$ is strictly zero. This actually
holds  beyond  the semiclassical approximation, as it  can be traced back to
conservation of probability by the master equation for $k=0$
\cite{PBraun97}.    
\item $R$ is an even function of $\eta$ and has in fact always a {\em
maximum} at $\eta=0$. 
\end{itemize}

An equation like (\ref{HJ}) has already been found for a dissipative problem in
\cite{Kitahara75}, 
but the preexponential factor which is crucial to our problem was not
treated there. That prefactor is obtained in next-to-leading order in 
$1/J$ as \cite{PBraun97.2} 
\begin{eqnarray} \label{eq:B}
B(\mu,\nu;\eta;s)&=&\sqrt{\frac{J}{2\pi
}}\sqrt{\left.\frac{\partial \nu}{\partial \mu}\right|_{a}}\sqrt{\dm\dn R(\mu,\nu;\eta;s)}\,;
\end{eqnarray}
it differs from the familiar form 
$\sqrt{\dm\dn\Sn}$ in the van Vleck
formula for unitary propagators by the additional factor 
$\sqrt{\frac{\partial \nu}{\partial \mu}\left.\right|_{a}}$.
Since $\sqrt{\dm\dn\Sn}$ arises from the propagation of a wave--function we
might expect  the square of this factor when propagating a density matrix
with $\eta=0$ (i.e.~probabilities). And indeed,
both square roots  in (\ref{eq:B}) are the same in the
semiclassical limit and combine to the classical Jacobi determinant: For
$J\to\infty$ we can restrict our attention to values of $\mu$ and $\nu$
close to the classical trajectory, i.e.~$a=1$, as the propagator for other
values is 
exponentially small. We call this trajectory
$\mu_d(\nu)$ where the time is fixed. The first square root
just gives 
$\sqrt{\frac{d\mu_d^{-1}(\mu)}{d\mu}}$.  On the
other hand, since $R(\mu,\nu;0;s)$ has a 
maximum on the classical trajectory, it can be written as
$R(\mu,\nu;0;s)\simeq-\frac{\alpha}{2}(\nu-\mu_d^{-1}(\mu))^2$ close to the
maximum with some  
parameter $\alpha>0$. This means  $\dm\dn
R(\mu,\nu;0;s)=\alpha\frac{d\mu_d^{-1}}{d\mu}$, and for $J\to\infty$ 
(\ref{eq:B}) therefore leads to the classical propagator of {\em
probabilities}, 
\[
D_{cl}(\mu,\nu,\;0;s)=\left|\frac{d\mu_d^{-1}}{d\mu}\right|
\delta(\nu-\mu_d^{-1}(\mu))\,. 
\]
As a conclusion, we may write the preexponential factor for $\eta=0$ as  
\begin{equation} \label{eq:dscl3}
B(\mu,\nu;0;s)\simeq\sqrt{\frac{J}{2\pi
}}\sqrt{\left|\frac{d\mu_d^{-1}(\mu)}{d\mu}\right|}\sqrt{\dm\dn
R(\mu,\nu;0;s)}\,,
\end{equation}
{\em if it is evaluated on the classical trajectory}. This is what
we will need for the 
semiclassical evaluation of the traces of $P$ (see the next
section).

A word is in order here about the limits of validity of
the present semiclassical treatment of dissipation.
It  disregards the existence of discrete energy levels and becomes
unacceptable when this discreteness is important. These are the cases when 
\begin{itemize}
\item{ the initial state $n$ and the final state $m$ coincide or are
separated by just a few intermediate levels: $n-m\sim 1$;} 
\item{ $m$ or $n$ (or both)  coincide with the highest or lowest energy
state of the system $|\pm j\rangle$ or are separated from them by just a few
levels, i.e. when at least one of the following relations hold  $\pm j\pm n
\sim 1, \pm j\pm m \sim 1$. } 
\end{itemize}

The first limitation implicitly puts a lower bound on the time, $s\gg {\cal
O}(J^{-1})$, because for small propagation times it is only the 
elements $D_{mn}$ with $n-m$ zero or of order unity which are
noticeably different from zero. The failure of the continuum approximation
is self evident in this case: the probability literally had no time to get
away from the initial level to form a smooth distribution. 
The second limitation means that the initial and final 
coordinates should not be too close to the ``poles'' $\pm j$. The
inapplicability of the 
naive semiclassical approximation to describe  the lowest energy state of a
quantum system is well-known; in  systems with the spectrum bounded from
above this is also true about the highest-energy state. Again, there is an
implicit limitation on the time which should be not too large (small compared
with $J$) so that the probability is not concentrated close to the ground
state. 

In the present work the deficiencies of the semiclassical
approximation just mentioned seem to be of no importance, at least for the
lowest traces of $P$. For higher traces we see deviations to next order in
$1/J$ which may well be due to our neglecting
the discreteness of the spectrum. A better
approximation for the dissipative propagator exists 
which allows for  uniform evaluation, at least for not very
large times \cite{PBraun97.2}. 

Since both $F$ and $D$ have a classical counterpart,
so does the combined map $P$. We shall denote it as $P_{cl}$ and propose to
show that its periodic 
points determine the spectral properties of $P$.

\section{Semiclassical evaluation of the first trace of $P$}\label{secttrP} 
We first write out an exact formal expression
for the quantum map $P$ in terms of the discrete quantum
numbers. Immediately after 
the unitary motion, $\rho$ has the elements $\rho_n(k,0+)=\langle
n_1|F\rho(0)F^{\dagger}|n_2\rangle=\sum_{l_1,l_2=-j}^jF_{n_1l_1}(F^{\dagger})_{l_2n_2}\langle
l_1|\rho(0)|l_2\rangle=\sum_{l,r}F_{n+k,l+r}F_{n-k,l-r}^*\rho_l(r,0)$ with
$l=\frac{l_1+l_2}{2}$ and $r=\frac{l_1-l_2}{2}$. We insert this in the
general solution (\ref{eq:rhoDrho}) and get 
$\rho_m(k,\tau)=\sum_{n,l,r}D_{mn}(k,\tau)F_{n+k,l+r}F_{n-k,l-r}^*\rho_l(r,0)
=\sum_{l,r}P_{mk;lr}\rho_l(r,0)$
where we have introduced the dimensionless parameter $\tau=2J\kappa T$. 
The propagator $P$ is read off as
\begin{equation} \label{eq:Pmklr}
P_{mk;lr}=\sum_nD_{mn}(k,\tau)F_{n+k,l+r}F_{n-k,l-r}^*,
\end{equation}
and yields the first trace
\begin{equation} \label{eq:trP1}
\tr P=\sum_{m,k}P_{mk;mk}=\sum_{m,n,k}D_{mn}(k,\tau)F_{n+k,m+k}F_{n-k,m-k}^*\,.
\end{equation}
\\
We now replace sums by integrals via
Poisson summation, employ the semiclassical propagators derived in the
previous section, and shall eventually integrate 
by the saddle--point method. The integrals to be done read
\begin{eqnarray}\label{spi}
\tr P&=&J^2\sum_{r,l,t=-\infty}^\infty \int 2d\mu d\nu
d\eta\exp[J\Phi(\mu,\nu;\eta)] 
\sum_{\sigma_1,\sigma_2}\re^{\ri\frac{\pi}{4}(\sigma_1+\sigma_2)}\nonumber\\
&&\cdot B(\mu,\nu,\eta)C(\nu+\eta,\mu+\eta)C^*(\nu-\eta,\mu-\eta)\,.
\end{eqnarray}
They involve the now complex action 
\begin{eqnarray}\label{eq:F} 
\Phi(\mu,\nu;\eta)&=&R(\mu,\nu;\eta)+\ri2\pi(r(\mu+\eta)+l(\nu+\eta)+2t\eta)\nonumber\\
&&+\ri\left\{
\sigma_1S(\nu+\eta,\mu+\eta)+\sigma_2S(\nu-\eta,\mu-\eta)\right\}\nonumber\\
&&=\Phi_0(\mu,\nu;\eta)+\ri\Phi_1(\mu,\nu;\eta)\,
\end{eqnarray}
whose real part is related to dissipation while the imaginary part
represents unitary dynamics (We apologize to the reader for frivolously
calling action what is usually called $-{\rm i}\times$ action). 
The integral contains factors $2$ in front of the integer
$t$ and before $d\mu$ which arise because $m$, $n$ and $k$  can be
simultaneously  half--integer. 
For brevity, we have suppressed the parameters $\beta$ and $\tau$. \\
Three complex  saddle--point equations (SPE) have to be fullfilled: $\dm
\Phi=0$, $\dn \Phi=0$ and $\de \Phi=0$. We look for real  solutions
(complexity would indeed look unphysical, as $m=\mu J \in {\cal
Z}$ etc.). Formally we can not exclude the existence of further complex
solutions, but as long as classical solutions exist we expect them to
dominate over non--classical ones. This is certainly true in the case of the
superradiance
dissipation where $R=0$ on the classical trajectory, whereas complex
solutions would lead to exponential supression from the term $i\Phi_1$. But
even more generally we expect classical solutions to dominate, since they
are known to dominate in non--dissipative quantum mechanics and dissipation
favors classical behavior even more. \\
We  thus need separately $\dm \Phi_0=0$
and $\dm \Phi_1=0$, where 
$\Phi_0$ and $\ri 
\Phi_1$ are the real and imaginary parts of $\Phi$ (and correspondingly for the
derivatives $\dn$ and $\de$). These are six equations for the three
variables $\mu, \nu, \eta$, but the 
additional three integers $r,l,t$ from the Poisson summation do allow for a
solution.\\ 
The first equation, $\de \Phi_0=0$, immediately gives $\eta=0$ with the
important consequence that only 
the propagation of 
probabilities contributes to the trace in the semiclassical limit.
Inserting $\eta=0$ in $\de \Phi_1=0$, we are led to
$(\sigma_1-\sigma_2)(\dn+\dm)\Sn+2\pi(r+l+2t)=0$. 

Consider first the case
$\sigma_1=-\sigma_2$. In view of the generating properties of $\Sn$,the
foregoing equation yields
$\phi_f(\nu,\mu)=\phi_i(\nu,\mu)+\pi(r+l+2t)$. From $\dm 
\Phi_1=0=2\pi r+(\sigma_1+\sigma_2)\dm \Sn=0$ we conclude $r=0$; and from $\dn
\Phi_1=0$  correspondingly $l=0$. We thus obtain the  unique classical
trajectory for the unitary evolution which connects the initial and final
momenta for the given time span $\tau$, uniqueness due to the additional
requirement that initial and final azimuths  be the same up 
to a multiple of $2\pi$! The second of the three saddle-point equations for
the real part,  $\dn \Phi_0=0$, 
immediately leads to $a=1$ (since $\eta=0$), i.e. precicely the condition
distinguishing the trajectories of the classical overdamped pendulum  
within the larger family of Hamiltonian tracjectories describing the
semiclassical limit of the dissipative motion.  
With that condition met, the last equation 
$0=\dm \Phi_0=-\dn \Phi_0$ also holds without further restriction.  \\
To summarize, the saddle-point equations determine the classical trajectory
which leads, in the unitary
part of the map, from $\mu$ to $\nu$ with  constant $\phi$ and,
for the 
dissipative part, back from $\nu$ to $\mu$, again under constant $\phi$. The
saddle points of the above integrand are the {\em fixed  
points of the classical map}. \\

Consider now $\sigma_1=\sigma_2\equiv\sigma$. The imaginary part of the
$\eta$ variation,  $\de \Phi_1=0$, does not  
restrict $\nu,\mu,\eta$ but just gives $r+l+2t=0$. New information now comes
from $\dn 
\Phi_1=0$ 
and $\dm \Phi_2=0$ and amounts to $\phi_f=\pi l=-\pi(r+2t)$ and $\phi_i=-\pi
r$. So
we have again $\phi_f=\phi_i$ up to an integer multiple of $2\pi$, but
additionally $\phi_i=0$ or $\pi$. The dissipative part remains unchanged,
such that for the combined map we get the particular class of periodic
orbits that take place at $\phi=0$ or $ \pi$. For our integrable dissipative
kicked top, such orbits are only possible
if the  
damping constant $\tau$ exceeds a certain critical value $\tau_c$. Moreover, we
shall see that these orbits lie exactly on the boundary between classically
allowed and classically forbidden motion concerning the rotation alone
(cf.~sec\ref{secuni}). The 
semiclassical approximation (\ref{eq:dmn})  then ceases to be valid and
should be replaced by 
a uniform approximation bridging the classically allowed
and forbidden domains \cite{PBraun96}. This program is beyond the scope of
the present work; we will restrict ourselves to regimes where
this particular kind of orbits does not exist. 

Note that the combination $\sigma_1=-\sigma_2$ would arise
automatically, if the van Vleck propagator contained only one classical 
path. The
opposite sign results from  the time reversed propagator $F^{\dagger}$. It
means 
the interference of a given 
orbit with itself. \\

The fact that only the interference of an orbit with itself contributes can
be seen 
more generally in the case where the unitary part contains several classical
paths
with actions $S_l(\nu,\mu)$. One encounters then a term $\ri J
(S_l(\nu+\eta,\mu+\eta)-S_k(\nu-\eta,\mu-\eta))$ and a double sum over paths
$l,k$. However, the SPEs lead to an equality of initial and final angles
(modulo $2\pi$) for the two paths $l$ and $k$.
Since to a given initial
phase--space point there is only one  classical path, 
only $l=k$ can contribute. Even the case of $l$ and $k$ designating two
branches of a  separatrix in phase space can
in general be excluded, due to the equality also of the final
phase--space 
point. The only exception are degenerate 
paths by which we mean that two originally separate paths conincide when a
system parameter is suitably changed. Exactly this extraordinary situation
occurs in the above 
example for the kicked top: For $\tau<\tau_c$ there are always two symmetry
related classical
paths for the unitary part (parametrized by $\sigma=\pm$), one starting at
an azimuth $\phi_i$, the other at an azimuth $-\phi_i$. When $\tau$ is
increased 
both conincide for $\tau=\tau_c$ at $\phi_i=0$ and remain there for all values
$\tau>\tau_c$. But at the same time classical dynamics breaks down and the
van Vleck propagator ceases to be valid, 
anyway.    

To exploit the SPA, we have to expand $\Phi$ to second order in the
deviation $\bx^T=(\delta\mu,\delta\nu,\delta\eta)$ from  the
fixed point $(\mu_0,\nu_0;0)$. We obtain a quadratic form
with a complex symmetric $3\times 3$ matrix $Q_{3\times 3}$ that contains
the second partial derivatives
$\partial_{x_i}\partial_{x_j}\Phi(\mu,\nu;\eta)\left.\right|_{\mu_0,\nu_0;0}$
in an obvious way,
\begin{equation} \label{eq:FQ}
\Phi(\mu,\nu;\eta)\simeq \Phi(\mu_0,\nu_0;0)-\frac{1}{2}\bx^TQ_{3\times
3}\bx\,.
\end{equation}
We have chosen to pull out a minus sign from the partial derivatives.
The fact that $Q_{3\times 3}$ is complex and therefore non--Hermitian makes
the SPA non--trivial. 
In view of the repeated application of the saddle--point method for the 
integration of a complex function of several variables, we devote appendix
\ref{appSaddle} to a discussion of 
the general formula derived in \cite{Fedoryuk87,Prudkovsky74}. 

Let us come back to our saddle-point integral (\ref{spi}). According to
equation (\ref{eq:SPA}) we  have to evaluate $\det Q_{3\times
3}$. 
It is easy to see that the $2\times 2$ ($\mu,\nu$) block and the $(\eta,\eta)$
block of the imaginary part of $\Q{3}$ are zero. In addition, $\dm\de \Phi_0=0=\dn\de
\Phi_0$, since $\Phi_0$ has a maximum at $\eta=0$ for all $\mu$ and $\nu$,
i.e.~it
must have the structure $\Phi_0\simeq f_0(\mu,\nu)\eta^2$ for small $\eta$ with
some function $f_0$.
The second derivatives of the real part  can be simplified on the classical
periodic orbit: 
Since 
$\dm \Phi_0(\mu,\nu;0)\left.\right|_{\mu=\mu_d(\nu)}=0$ for {\em all} $\nu$,
we find 
$\frac{d}{d \nu}\dm
\Phi_0(\mu,\nu;0)|_{\mu=\mu_d(\nu)}=\dm^2\Phi_0\frac{d\mu_d}{d\nu}+\dm\dn \Phi_0=0$,
i.e.~$\dm\dn \Phi_0=-\dm^2\Phi_0\frac{d\mu_d}{d\nu}$. Similarly, starting from
$\dn \Phi_0 \left.\right|_{\mu=\mu_d(\nu)}=0$, one finds $\dn^2
\Phi_0=\dm^2\Phi_0\left(\frac{d\mu_d}{d\nu}\right)^2$. We introduce the
functions 
$h(\nu,\mu)=2(\dn+\dm)\Sn=\de \Phi_1$ and $\gamma=\frac{d\mu_d}{d\nu}$ to
write $\Q{3}$ in the form
\begin{equation} \label{eq:Q3}
\Q{3}=-\left(\begin{array}{ccc}
\dm^2\Phi_0&-\gamma \dm^2\Phi_0 &\ri\dm h\\
-\gamma \dm^2\Phi_0&\gamma^2 \dm^2\Phi_0&\ri\dn h\\
\ri\dm h& \ri\dn h& \de^2\Phi_0
\end{array}
\right)\,.
\end{equation} 
It has the determinant 
\begin{equation} \label{eq:detQ3}
\det\Q{3}=-\dm^2\Phi_0(\dn h+\gamma\dm h)^2\,.
\end{equation}
Since at the saddle point $\dm^2\Phi_0<0$ (maximum of the propagator on the
classical trajectory!), we have $\det\Q{3}>0$ and by the same reason the
first minor $D_1=-\dm^2\Phi_0>0$. The second upper left minor gives zero,
but we can fix this by introducing a small positive $\epsilon$ in the first
element. This leads to $D_2=-\epsilon\gamma^2\dm^2\Phi_0>0$. Thus, all
minors are real and positive so that $\arg\det\Q{3}$ is now uniquely fixed
to zero (cf.~(\ref{arg})).\\   
The preexponential factor in $D$ simplifies on 
the classical trajectory (cf.~(\ref{eq:dscl3}): 
$\sqrt{\left|\frac{d\mu_d^{-1}(\mu)}{d\mu}\right|}\sqrt{\dm\dn
R(\mu,\nu;0)}=\sqrt{\frac{d\mu_d^{-1}(\mu)}{d\mu}}\sqrt{-\frac{d\mu_d}{d\nu}\dm^2\Phi_0}=\sqrt{-\dm^2\Phi_0}$.
From the action in the exponent only the real part $R=R(\mu,\nu;0)$ remains.
We arrive at the saddle--point approximation for $\tr P$, 
\begin{equation} \label{eq:trP1S}
\tr P=\sum_{p.o.}\left|\frac{\dm\dn\Sn}{(\dn^2+\dm\dn)\Sn+\frac{d \mu_d}{d\nu}(\dm^2+\dm\dn)\Sn}\right|\re^{JR}\,.
\end{equation}
The sum is over periodic orbits of the classical map and it is understood
that all quantities are evaluated on these.\\  
The foregoing trace formula can be 
simplified further if we go over to phase--space coordinates $\mu$ and
$\phi=\phi_i$ and exploit the generating properties of $\Sn$.  Let us call
$\nu_r(\mu,\phi)$ the momentum component 
of the classical trajectory for the 
unitary motion (the index $_r$ stands for rotation), if the initial values 
$\mu$ and $\phi$ are given. The
corresponding angular coordinate will be denoted by
$\phi_r(\mu,\phi)$; it is related to $\phi_f(\nu,\mu)$ by
$\phi_f(\nu,\mu)=\phi_r(\mu,\phi_i(\nu,\mu))$. The 
corresponding quantities for the dissipative motion are
$\mu_d(\nu,\phi)=\mu_d(\nu)$  and $\phi_d (\nu,\phi)=\phi$. The generating
properties of $\Sn$ (cf.~(\ref{eq:gen})) imply
 $\frac{\dm^2\Sn}{\dm\dn\Sn}=\frac{\dm \phi_i(\nu,\mu)}{\dn
\phi_i(\nu,\mu)}=-\frac{\partial\nu_r(\mu,\phi)}{\partial\mu}\left.\right|_{\phi=\phi_i}$ and
$\frac{\dn^2\Sn}{\dm\dn\Sn}=-\frac{\dn\phi_f(\nu,\mu)}{\dn\phi_i(\nu,\mu)}=-\frac{\partial\phi_r(\mu,\phi)}{\partial
\phi}$.
Our trace thus reads
\begin{equation} \label{eq:trP1S2}
\tr
P=\sum_{p.o.}\frac{\re^{JR}}{\left|1+\frac{d\mu_d}{d\nu}-\frac{\partial
\phi_r(\mu,\phi)}{\partial\phi}-\frac{d\mu_d}{d\nu}\frac{\partial \nu_r(\mu,\phi)}{\dm}\right|}\,.
\end{equation}
Finally, the preexponential factor can be written in terms of traces
of classical monodromy matrices. The monodromy matrices for the
dissipative and the unitary motion are 
\begin{equation}
M_d=\left(
\begin{array}{cc}
\left(\frac{\partial \mu_d}{\partial \nu}\right)_\phi&0\\
0&\left(\frac{\partial \phi_d}{\partial \phi}\right)_\mu
\end{array}
\right)\mbox{, }\quad
M_r=\left(
\begin{array}{cc}
\left(\frac{\partial \nu_r}{\partial \mu}\right)_\phi&\left(\frac{\partial \nu_r}{\partial \phi}\right)_\mu\\
\left(\frac{\partial \phi_r}{\partial \mu}\right)_\phi&\left(\frac{\partial \phi_r}{\partial \phi}\right)_\mu
\end{array}
\right)\,.
\end{equation}
respectively.  Their product $M=M_dM_r$  is the monodromy
matrix for the combined map. 
Recall that the dissipation conserves $\phi$, such that
$\frac{d\mu_d}{d\nu}=\left(\frac{\partial\mu_d}{\partial\nu}\right)|_\phi$.
We therefore arrive at
\begin{equation} \label{eq:trP1fin}
\tr P=\sum_{p.o.}\frac{\re^{JR}}{\left|\tr M_d-\tr M\right|}\,.
\end{equation}
This is the first central result. It generalizes Tabor's
formula for classically area--preserving maps to an area--nonpreserving map
and shows that even in the
case of dissipative quantum maps, information about the spectrum is encoded
in classical periodic orbits. All quantities must be evaluated on the
periodic orbits and with $\eta=0$.
We remark that the formula holds for both chaotic or
integrable maps, as long as the periodic orbits are
sufficiently well separated in phase space such that the SPA is valid.
\\
For comparing (\ref{eq:trP1fin}) with Tabor's result, $\tr
F=\sum_{p.o.}\frac{\re^{\ri(JS+\alpha)}}{|2-\tr M|^{1/2}}$, one should
remember that we consider the propagator of the density matrix, but Tabor the
propagator of the wave function \cite{Tabor83}. In the limit of zero
dissipation, we should 
get $\tr P=|\tr F|^2$. That limit can unfortunately not be taken in
(\ref{eq:trP1fin}), since our semiclassical dissipative
propagator is only valid for $\tau\gtrsim 1/J$. However, 
$|\tr F|^2$ for $\tau\to0$ would definitely lead to a double sum over periodic
orbits. Of the double sum only a single sum
remains in (\ref{eq:trP1fin}); all cross terms between different orbits
are killed by 
dissipation (this is due to $\eta=0$!). An important
quantum--mechanical effect of the dissipation is indeed the very rapid
destruction of interferences on time scales much shorter than the classical
time scales \cite{Haake91}. {\em Decoherence  leads in the
trace automatically to the  ``diagonal approximation''} which suppresses
interferences of different orbits!\\
Tabor's preexponential factor is reproduced in the limit $\tau\to 0$, since
$M_d$ becomes the unit matrix, and thus $\tr M_d=2$ and $\tr M=\tr
(M_dM_r)=\tr 
M_r$. It is, however, raised to the power 1 instead of $1/2$, since we
propagate a 
density matrix and not a wave function. The 
action itself is purely real; the imaginary parts from the unitary motion
cancel each other at $\eta=0$. This property holds for all dissipative
processes describable by a Markovian master equation that leads to a
dissipative propagator with a single maximum (as a function of $\eta$) at
$\eta=0$. 
As outlined  in the section on
the dissipative part, the saddle-point action is zero in our
case of spin damping. Then all physics is
in the preexponential factor, which describes the  stability of each orbit!
If $R\ne 0$, only a single periodic orbit may dominate the trace in the
limit of $J\to\infty$, namely the one with the maximum value of $R$.

Before we calculate higher traces, let us comment about the generality of
our result. In deriving (\ref{eq:trP1fin})  we made use of the 
general van
Vleck forms  of the propagators 
and the generating properties of the actions appearing in them. Within the
class of dissipative processes considered
the only 
ingredients particular to the  specific problem at hand were the fact that the
dissipation conserves $\phi$ and the sum over two
paths in (\ref{eq:dmn}). However, since the dissipative processes considered
impliy only interference of each orbit with itself, the latter feature
does not restrict the generality of 
(\ref{eq:trP1fin}). 
The formula should hold independently of the number of classical paths that
contribute to the unitary motion. The vanishing of the Maslov phase relies
only on  
properties of the dissipative part and will always hold if the
dissipative propagator has a maximum on the classical trajectory.

\section{Higher traces}\label{secttrPn}
We will now show how the semiclassical trace formula (\ref{eq:trP1fin})
generalizes to higher traces $\tr 
P^N$. For our  $2j+1$ dimensional Hilbert space, the knowledge of $(2j+1)^2$
traces  
suffices in principle to reconstruct the entire spectrum. As a first step
we look at the second trace.
\subsection{The second trace}\label{secttrP2}
In principle, $\tr P^2$ could be calculated just as $\tr P$ by writing
down the exact discrete sums as in (\ref{eq:trP1}), transforming them to
integrals, and doing the integrals by the SPA. The quadratic form in the
exponent 
would then give rise to a $6 \times 6$ matrix, and in general, for the $N$--th
trace, to a $3N\times 3N$ matrix. It turns out that the problem 
simplifies considerably if one integrates over $\nu$ once for all in the
propagator 
itself, so that the latter will only depend on actual initial and final
coordinates. The matrix in the quadratic form arising from the SPA is then
only a $2N\times 2N$ matrix. \\
Let us therefore write an arbitrary matrix element of $P$ as  
\begin{eqnarray}
P_{mk;m'k'}&=&\sum_{l=-\infty}^\infty\int
d\nu\sum_{\sigma_1,\sigma_2}B(\mu,\nu,\eta;\tau)C(\nu+\eta,\mu'+\eta')\nonumber\\
&&\cdot C^*(\nu-\eta,\mu'-\eta')\exp\left(JG(\mu,\eta;\mu',\eta';\nu)\right)\,,
\end{eqnarray}
where again $m=\mu J$ etc.~and $C(\mu,\nu)$ is defined in
(\ref{CNM}). The complex action $G$ reads 
\begin{equation} \label{eq:G}
G(\mu,\eta;\mu',\eta';\nu)=R(\mu,\nu;\eta,\tau)+\ri
\sigma_1S(\nu+\eta,\mu'+\eta')+\ri\sigma_2S(\nu-\eta,\mu'-\eta')+\ri 
2\pi
l\nu\,.
\end{equation}
The saddle--point equation 
\begin{equation} \label{eq:dG}
\dn G=0=\dn R+\ri (
\sigma_1\dn S(\nu+\eta,\mu'+\eta')+\sigma_2\dn S(\nu-\eta,\mu'-\eta')+2\pi l)
\end{equation}
will in general have  complex solutions
$\nu=\overline{\nu}(\mu,\eta;\mu',\eta')$, as long as $\eta$ and $\eta'$
are not zero 
and $\mu$ and $\mu'$ are not connected by a classical trajectory. It may
even be 
possible that several solutions or no solution at all exist. However, note
that the saddle-point equation (\ref{eq:dG}) simplifies considerably and
gives a physically 
meaningful fixed point if $\sigma_1=-\sigma_2$ and $\eta=\eta'=l=0$ (and
that is the situation we encounter for our dissipative integrable
kicked top):  We then have
$\dn R(\mu,\nu;\eta,\tau)=0$ which is equivalent to the classical
dissipative equation of motion $\nub=\mu_d^{-1}(\mu)$ and $a=1$. \\ 
When the $\nu$ integral is done by a SPA the second derivative $\partial^2_\nu
G|_{\nu=\nub}\equiv\partial^2_\nub G$ comes into play; it can be combined with the
other preexponential factors of $P$. 
One needs a relation between second
derivatives of $G$ which can be obtained by differentiating  
with respect to $\mu$ in the saddle-point equation (\ref{eq:dG}), accounting for the $\mu$ dependence of $\nub$, 
\begin{equation} \label{d2G}
\partial^2_\nub G=-\dm \partial_\nub G\frac{1}{\frac{\partial \nub}{\partial
\mu}}=-\dm \partial_\nub R(\mu,\nub;\eta,\tau)\frac{1}{\frac{\partial \nub}{\partial
\mu}}\,.
\end{equation}
With the  abbreviation
$\Gb(\mu,\eta;\mu',\eta')=G(\mu,\eta;\mu',\eta';\nub(\mu,\eta;\mu',\eta'))$
for the action  at the saddle point we find
\begin{eqnarray}
P_{mk;m'k'}&=&\sum_{l=-\infty}^{\infty}\sum_{\sigma_1,\sigma_2}\sum_{\nub}\sqrt{\frac{\partial
\nub}{\partial \mu}}\sqrt{\left.\frac{\partial \nub}{\partial \mu}\right|_a}\label{P2t2}\nonumber\\
&&\cdot C(\nub+\eta,\mu'+\eta')C^*(\nub-\eta,\mu'-\eta')\exp\left(J\Gb(\mu,\eta;\mu',\eta')\right)\,.
\end{eqnarray}
The sum
over $\nub$ picks up all relevant saddles. To avoid possible confusion we
should add that of the two derivatives $\frac{\partial \nub}{\partial \mu}$
only the one originating from the semiclassical dissipative propagator is
meant at constant $a$, cf. (\ref{eq:B}), as long as $\mu$ and $\mu'$ are not
connected by the classical dissipative trajectory.

We are now ready to apply the map twice. Let us call  $R^i$ and $D^i$ the
 $i$-th unitary and dissipative step of the iterated map. The coordinates $(\mu,\eta)$ are transformed in
the following way:
\[
(\mu_1,\eta_1)\stackrel{R^1}{\longrightarrow}(\nu_1,\eta_2)\stackrel{D^1}{\longrightarrow}(\mu_2,\eta_2)\stackrel{R^2}{\longrightarrow}(\nu_2,\eta_3)\stackrel{D^2}{\longrightarrow}(\mu_3,\eta_3)=(\mu_1,\eta_1)\,.
\]
The last equation is the  periodicity condition for $\tr P^2$. The total
action  $\Psi_2$ that appears in the expression for $\tr P^2$ is
basically the sum of two actions 
$\Phi$ (cf.~(\ref{eq:F})), one from each repetition of the map. All
variables carry an additional index $i=1,2$ counting the
iteration of the map (in the case of the $\sigma$ it is the first
index),
\begin{eqnarray}
\Gb_2&=&R(\mu_1,\nub_2;\eta_1)+R(\mu_2,\nub_1;\eta_2)+\ri\sigma_{21}S(\nub_2+\eta_1,\mu_2+\eta_2)\nonumber\\
&&+\ri\sigma_{22}S(\nub_2-\eta_1,\mu_2-\eta_2)+\ri\sigma_{11}S(\nub_1+\eta_2,\mu_1+\eta_1)+\ri\sigma_{12}S(\nub_1-\eta_2,\mu_1-\eta_1)\nonumber\\
&&+\ri 2\pi((r_1(\mu_1+\eta_1)+l_1(\nub_2+\eta_1)+2t_1\eta_1+r_2(\mu_2+\eta_2)+l_2(\nub_1+\eta_2)+2t_2\eta_2)\nonumber\,.
\end{eqnarray}
The last line comes from the Poisson summation over $r_i,l_i$, and
$t_i$. The $\mu$'s and $\eta$'s 
are the independent integration variables whose values will now be
determined by 
saddle--point equations. \\
We assume again that the solution
of the SPE's is real and split them into their real and imaginary
parts. From the real part of the SPE's with respect to $\eta$,
$\dE{1}\Gb_2=0=\dE{2}\Gb_2$, we get  $\eta_1=\eta_2=0$. 
The real part of $\dM{1}\Gb_2=\dM{2}\Gb_2=0$ leads
to the 
classical dissipative equations of motion at constant $\phi$.  The imaginary
parts of $\dE{1}\Gb_2=0=\dE{2}\Gb_2$ give
\begin{eqnarray} \label{imdG}
(\sigma_{21}-\sigma_{22})\phi_f(\nub_2,\mu_2)&=&(\sigma_{11}-\sigma_{12})\phi_i(\nub_1,\mu_1)+(r_1+l_1+2t_1)2\pi\\
(\sigma_{21}-\sigma_{22})\phi_i(\nub_2,\mu_2)&=&(\sigma_{11}-\sigma_{12})\phi_f(\nub_1,\mu_1)+(r_2+l_2+2t_2)2\pi\,.
\end{eqnarray}
These are conditions on the initial and final azimuthal angles for the two
classical maps corresponding  to the two unitary parts. Let us look at them
more closely. Suppose that in
any of these equations $\sigma_{i1}=\sigma_{i2}$. We would get 
$\phi_i=0$ or $\phi_f=0$, and we have excluded such cases from our present
analysis. If on the other hand we had
$(\sigma_{21}-\sigma_{22})=-(\sigma_{11}-\sigma_{12})\ne 0$, this
would imply $\phi_f(\nub_1,\mu_1)=-\phi_i(\nub_2,\mu_2)$ (modulo $\pi$),
which is not 
possible since  during dissipation $\phi$ is conserved. We therefore conclude
 that only the two combinations
$\sigma_{i1}=-\sigma_{i2}\equiv\sigma=\pm 1$ contribute. Then the SPE's
define a period-2 point of the
combined map.\\
If we expand
$\Gb_2$ around a periodic point into a quadratic form,  a $4\times 4$
matrix $Q_{4\times 4}$ of second derivatives appears which consists of
$2\times 2$  blocks $(\mu,\eta)$, $(\eta,\mu)$, and $(\eta,\eta)$ where
$(\mu,\eta)$ contains the 
mixed derivatives 
$-\dmi\dej\Gb_2$ (and correspondingly for the other blocks):
\begin{equation} \label{Q4}
Q_{4\times 4}=\left(
\begin{array}{cc}0&(\mu,\eta)\\
(\eta,\mu)&(\eta,\eta)
\end{array}
\right)\,.
\end{equation}
The $(\mu,\mu)$
block vanishes. To see this, observe first that its imaginary part gives
zero as  $\sigma_{i1}=-\sigma_{i2}$. For the real part we have  by
construction $\dNb{2} \Gb_2=0=\dNb{2}R(\mu_1,\nub_2;0)$ {\em for all}
$\mu_1$. But then also $\dM{1}R(\mu_1,\nub_2;0)=0$ holds {\em for
all} $\mu_1$ (cf.~the general properties of $R$!). This
immediately gives
$\dM{1}\dM{1}\Gb_2=0$. A corresponding argument holds for
$\dM{2}\dM{2}\Gb_2$.  Furthermore, as $\nub_2=\mu_d^{-1}(\mu_1)$ on the
classical trajectory, $\mu_1$ does not even appear in $\dM{1}
R(\mu_1,\nub_2;0)$  and 
we thus obtain $\dM{2}\dM{1} R(\mu_1,\nub_2;0)=0$.

Given the structure of the  matrix $Q_{4\times 4}$, the $(\eta,\eta)$ block
is irrelevant, since \cite{Gantmacher86}
\begin{equation} \label{detQ4}
\det \Q{4}=(\det(\mu,\eta))^2\,.
\end{equation}
In order to calculate the mixed derivatives in the $(\mu,\eta)$ block, one
needs 
partial derivatives of the $\nub$'s with respect to the $\eta$'s, as we have
for 
example 
\begin{equation} \label{dedmG}
\dE{2}\dM{1}\Gb_2=\partial_{\nub_2}\dM{1} R\frac{\partial\nub_2}{\partial\eta_2}+2\ri\sigma\dM{1}\dNb{1} S(\nub_1,\mu_1)\,;
\end{equation}
they are obtained by totally differentiating $\dNb{i} \Gb=0$ with respect to
the $\eta$'s and then setting $\eta_1=\eta_2=0$; the one needed  for the
above example reads
\begin{equation} \label{dnb2}
\frac{d\nub_2}{d\eta_2}\left.\right|_{\eta_1=\eta_2=0}=-2\ri\sigma\frac{\dNb{2}\dM{2} S(\nub_2,\mu_2)}{\partial^2_{\nub_2}R(\mu_1,\nub_2;0)}\,.
\end{equation}
Upon transforming the remaining second mixed derivative of $R$ in
(\ref{dedmG}) into second pure 
derivatives as in the case of $\tr P$ we can eliminate the derivatives of
$R$ entirely. We find for the above example
\begin{equation} 
\dE{2}\dM{1}\Gb_2=2\ri\sigma
\left(
\frac{\partial\nub_2}{\partial\mu_1}\dNb{2}\dM{2} S(\nub_2,\mu_2)+\dM{1}\dNb{1} S(\nub_1,\mu_1)
\right)\,.
\end{equation}
In a similar way the remaining three partial derivatives of $\Gb_2$ can be
found. 
If we abbreviate $S_i=S_i(\nub_i,\mu_i)$ ($i=1,2$,  $\nub_3\equiv\nub_1$) and
$R_i=R_i(\mu_i,\nub_{i+1};0)$), $\det
\Q{4}$ is given by 
\begin{eqnarray}
\det\Q{4}&=&4^2\left[\right.\frac{\partial\nub_2}{\partial\mu_1}\left((\dNb{2}^2S_2)(\dM{2}^2S_2)-(\dNb{2}\dM{2}
S_2)^2\right)+\frac{\partial\nub_1}{\partial\mu_2}\left((\dNb{1}^2S_1)(\dM{1}^2S_1)-(\dNb{1}\dM{1}
S_1)^2\right)\nonumber\\
&&+\frac{\partial\nub_2}{\partial\mu_1}\frac{\partial\nub_1}{\partial\mu_2}\left((\dNb{1}^2S_1)(\dNb{2}^2S_2)-(\dNb{1}\dM{1}
S_1)(\dNb{2}\dM{2} S_2)\right)\nonumber\\
&&+(\dM{1}^2S_1)(\dM{2}^2S_2)-(\dNb{1}\dM{1} S_1)(\dNb{2}\dM{2} S_2)
\left.\right]^2
\end{eqnarray}
and entails 
\begin{equation} \label{trP2}
\tr P^2=4\sum_{p.p.}\frac{\left| (\dNb{1}\dM{1} S_1)(\dNb{2}\dM{2} S_2)\frac{\partial\nub_1}{\partial\mu_2}\frac{\partial\nub_2}{\partial\mu_1} \right|}{\left|\det\Q{4}\right|^{1/2}}\re^{J(R_1+R_2)}\,.
\end{equation}
The sum is over all periodic points of the map $P^2$. The various prefactors
from 
the dissipative parts have canceled, up to two factors $\frac{\partial
\nub_i}{\partial \mu_{i+1}}=\frac{d\mu_d^{-1}}{d\mu_{i+1}}=1/\frac{d\mu_d(\nub_i)}{d\nub_i}$.
Again,
for each periodic point 
the overall sign of the summand should be determined. 
It turns out to be
positive (for a proof we refer to the discussion in the case of the $N$th
trace below).
The preexponential factor can again be
written in terms of elements of the classical monodromy matrices, as 
can be verified in a straightforward but  lengthy calculation. We
introduce monodromy matrices $M_d^{(i)}$ and $M_r^{(i)}$ for the 
dissipative and unitary evolution  in the $i$--th step,
respectively, and 
\begin{equation} \label{prodM}
M^{(i)}=M_d^{(i)}M_r^{(i)}
\end{equation}
for the entire $i$--th
step. The total monodromy matrix $M$ is given by the product 
$M=M^{(N)}\ldots M^{(1)}\equiv \prod_{i=N}^1 M^{(i)}$. With this
convention, the second trace reads
\begin{equation} \label{trP2fin}
\tr P^2=\sum_{p.p.}\frac{\re^{J(R_1+R_2)}}{\left|\tr M_d^{(2)}M_d^{(1)}-\tr
M\right|} \,.
\end{equation}

\subsection{The $N$--th trace}\label{secttrPN}
Our results for the first and the second trace suggest a 
generalization for the $N$--th trace,
\begin{equation} \label{trPN}
\tr P^N=\sum_{p.p.}\frac{\re^{J\sum_{i=1}^N R_i}}{\left|\tr \prod_{i=N}^1M_d^{(i)}-\tr
M\right|}\,,
\end{equation}
where the sum is now over periodic points of $P_{cl}^N$. The rest of this
section is devoted to the proof of the formula.\\
Starting point is  the intermediate semiclassical  form (\ref{P2t2}) of
the propagator $P$. We have to consider  a sequence 
of intermediate coordinates of the form
\begin{equation} \label{seq}
{\mu_1\choose\eta_1}\stackrel{R^1}{\longrightarrow}{\nu_1\choose\eta_2}\stackrel{D^1}{\longrightarrow}{\mu_2\choose\eta_2}\stackrel{R^2}{\longrightarrow}{\nu_2\choose\eta_3}\stackrel{D^2}{\longrightarrow}{\mu_3\choose\eta_3}\ldots\stackrel{R^N}{\longrightarrow}{\nu_N\choose\eta_{N+1}}\stackrel{D^N}{\longrightarrow}{\mu_{N+1}\choose\eta_{N+1}}={\mu_1\choose\eta_1}\,.
\end{equation}
All variables are periodically continued in their indices
($\nu_{N+i}\equiv \nu_i$, etc.). The total action is
given by
\begin{eqnarray}
\Gb_N&=&\sum_{i=1}^NR(\mu_i,\nub_{i-1};\eta_i)+\ri\sum_{i=1}^N\left(\sigma_{i1}S(\nub_i+\eta_{i+1},\mu_i+\eta_i)+\sigma_{i2}S(\nub_i-\eta_{i+1},\mu_i-\eta_i)\right)\,.
\end{eqnarray}
We have omitted for clarity the term coming from the Poisson summation which
is easily seen to lead to integer multiples of $2\pi$ in the 
equality of initial and final angles that we shall find presently. 
For convenience we introduce a vector notation for the sets of variables,
$\mbox{\boldmath $\mu$}=\mu_1,\ldots,\mu_N$, and similarly for {\boldmath $\nub$} and {\boldmath $\eta$}. The functions
$\nub_i=\nub_i(\mu_{i+1},\eta_{i+1};\mu_i,\eta_i)$ satisfy 
\begin{equation} \label{dnubG}
\dNb{i} \Gb_N(\mbox{\boldmath $\mu, \nub, \eta$})=0\,.
\end{equation}
Everything goes through for the saddle--point equations as for the first
and second trace.
Thus, $\sigma_{i1}=-\sigma_{i2}=\sigma=\pm 1$, 
and the SPE's pick up the periodic
points of $P_{cl}^N$. 
The $2N\times 2N$ matrix $\Q{2N}$ has the same block structure as
in the case $N=2$, that is $\dmi\dmj\Gb=0$ for all $i,j$. Therefore its
determinant depends only on  the $N\times N$ ($\mu,\eta$) block
which we will call $B_{N\times N}$ in the following \cite{Gantmacher86}: 
\begin{equation} \label{Q2N}
\det\Q{2N} =(-1)^N(\det B_{N\times N})^2\,.
\end{equation}
The partial derivatives in $B_{N\times N}$ are determined in the
same way as for $N=2$ and contain only derivatives
of the imaginary part of $\Gb$. We find
\begin{eqnarray} 
\partial_{\eta_k}\dM{l}\Gb_N&=&2\ri\sigma\Big(\dNb{l-1}\dM{l-1}S_{l-1}\delta_{l-1,k}+\left[\dNb{l-1}^2S_{l-1}\frac{\partial\nub_{l-1}}{\partial
\mu_l}+\partial_{\mu_l}^2S_l
\right]\delta_{l,k}\nonumber\\
&&+\dNb{l}\dM{l}S_{l}\frac{\partial\nub_{l}}{\partial
\mu_{l+1}}\delta_{l+1,k}\Big)\,,\label{klG}
\end{eqnarray}
where $\delta_{l,k}$ is the Kronecker--delta.\\

After the saddle--point integration is done, the expression for the $N$th
trace reads
\begin{equation} \label{trpn}
\tr P^N=2^N\sum_{p.p.}\left(\prod_{l=1}^N\frac{\partial \nub_l}{\partial
\mu}\big|_{a=1}|C(\nub_l,\mu_l)|^2\right)\sqrt{\frac{(2\pi)^{2N}}{J^{2N}|\det
B_{N\times N}|^2}}\re^{J\sum_{i=1}^NR_i}\,.
\end{equation} 
 To
proceed, one needs to calculate the determinant of
$B_{N\times N}$, which 
is a tridiagonal non--symmetric matrix with additional non-zero elements in
the upper right and lower left corners as 
well. The calculation of such a  determinant can be found in the appendix
\ref{appdet}, where we show that the determinant is the difference between
the traces of  two different products of $2\times 2$ matrices. Combining
equations 
(\ref{klG}), (\ref{finaldet}) and (\ref{trpn}) we are lead to 
\begin{equation} \label{trpn2}
\tr P^N=\sum_{p.p.}\frac{\re^{J\sum_{i=1}^NR_i}}{|\tr \prod_{l=N}^1
M_d^{(l)}-\tr \prod_{l=N}^1 M_l|}\,.
\end{equation} 
The inverted order of the indices at the product indicates that the matrices
are ordered from left to right according to decreasing indices. The matrix
$M_d^{(l)}$ 
in the denominator is already the monodromy matrix for the purely
dissipative part, $M_d^{(l)}={\rm
diag}(\frac{d\mu_d(\nub_l)}{d\nub_l},1)$. Unfortunately, a corresponding
statement 
does not hold for $M_l$  which is given by
\begin{equation} 
M_l=-\frac{1}{\partial_{\nu_l}\partial_{\mu_l} S_l}\left(
\begin{array}{cr}
\dnnS{l}+(\dmmS{l+1})\frac{d\mu_d(\nub_l)}{d\nub_l}&\mbox{ } -(\dnmS{l})^2\\
\frac{d\mu_d(\nub_l)}{d\nub_l}&0
\end{array}
\right)\,.
\end{equation}
Indeed,  the monodromy matrix     $M^{(l)}$ for the entire $l$th step
(cf. (\ref{prodM})) has the form
\begin{equation} 
M^{(l)}=\frac{1}{\partial_{\nu_l}\partial_{\mu_l} S_l}\left(\begin{array}{cc}
-(\dmmS{l})\frac{d\mu_d(\nub_l)}{d\nub_l}&\mbox{ } \frac{d\mu_d(\nub_l)}{d\nub_l}\\
-(\dnmS{l})^2+(\dnnS{l})(\dmmS{l})&-\dnnS{l}
\end{array}
\right)
\end{equation}
when expressed in terms of derivatives of $S_l$ and $\mu_d$. The fact that in
$M_l$ both the indices $l$ and $l+1$ appear makes it even impossible to find
a similarity transformation independent of $l$ that transforms $M_l$ to
$M^{(l)}$. Nevertheless, we will show now that the traces of $\prod_{l=N}^1
M_l$ and $\prod_{l=N}^1 M^{(l)}$ are equal. To do so, let us first pull out
the common factor $-1/\dnmS{l}$ from $M_l$ and $M^{(l)}$ by defining
$\overline{M}_l=(-\dnmS{l}) M_l$ and $\overline{M}^{(l)}=(-\dnmS{l})
M^{(l)}$. Then we separate the parts proportional to
$\frac{d\mu_d(\nub_l)}{d\nub_l}$ from  those independent of
$\frac{d\mu_d(\nub_l)}{d\nub_l}$, 
\begin{equation} \label{KL}
\ml=\frac{d\mu_d(\nub_l)}{d\nub_l}K_l+L_l\mbox{, and }\mL=\frac{d\mu_d(\nub_l)}{d\nub_l}K^{(l)}+L^{(l)}\,.
\end{equation}
We abbreviate 
$\alpha_l=\dmmS{l}$, $\beta_l=\dnnS{l}$ and $\gamma_l=\dnmS{l}$ and find that
the matrices on the rhs of the  equations in (\ref{KL}) have the structure
\begin{eqnarray}
K_l=\left(
\begin{array}{cc}
\alpha_{l+1}&0\\
1&0\end{array}
\right)
&\mbox{ , }&L_l=\left(
\begin{array}{cc}
\beta_l&-\gamma_l^2\\
0&0
\end{array}
\right)\\
K^{(l)}=\left(
\begin{array}{cc}
\alpha_l&-1\\
0&0
\end{array}\right)&\mbox{ , }&L^{(l)}=\left(
\begin{array}{cc}
0&0\\
\gamma_l^2-\alpha_l\beta_l&\beta_l
\end{array}
\right)\,.
\end{eqnarray}
Note that $\alpha_l$, $\beta_l$ and $\gamma_l$ are
periodic in the index, i.e.~$\alpha_{l+N}=\alpha_l$ etc.. The product
of the matrices $M_l$ (or $M^{(l)}$) gives rise to a sum of $2^N$ terms, each
of which can be written as some
factors $\frac{d\mu_d}{d\nub_l}$ multiplied with a trace over a product of
matrices $K_l$
and $L_l$ (or $K^{(l)}$  and $L^{(l)}$, respectively). To complete the
proof we must show that the traces over  products containing matrices $K_l$
and $L_l$ remain unchanged if we replace all $K_l$ by $K^{(l)}$ and
all  $L_l$ by $L^{(l)}$. Note that these products are always ordered
such that the indices decrease from left to right; and all indices from 1 to
$N$ appear.

We will show the equality of the traces  by
establishing rules  for calculating them  for all possible
combinations of $L$'s and $K$'s. The rules will turn out to be the same for
both upper and lower index $L$'s and $K$'s. 

Let us  express our matrices in terms of spin--$\frac{1}{2}$
operators $S_i=\hat{\sigma}_i/2$ and unity, where $i=x,y,z$ and the $\hat{\sigma}_i$ are
the Pauli matrices. In fact only raising and lowering
operators $S_\pm=S_x\pm iS_y$, and $\frac{1}{2}\pm S_z$ appear,
\begin{eqnarray}
K_l&=&S_-+\alpha_{l+1}(\frac{1}{2}+S_z)\\
L_l&=&-\gamma_l^2S_++\beta_l(\frac{1}{2}+S_z)\\
K^{(l)}&=&-S_++\alpha_l(\frac{1}{2}+S_z)\\
L^{(l)}&=&(\gamma_l^2-\alpha_l\beta_l)S_-+\beta_l(\frac{1}{2}-S_z)\,.
\end{eqnarray}
Such a decomposition greatly facilitates the calculation of products of
$K$'s and $L$'s due to the simple multiplication rules of the spin
operators. The rules can be summarized as
\begin{eqnarray}
S_-^2=S_+^2&=&0\mbox{,
}S_\pm S_\mp=(\frac{1}{2}\pm S_z) \\
(\frac{1}{2}\pm S_z)^n&=&(\frac{1}{2}\pm S_z)\mbox{ for all natural $n$;
}S_\pm(\frac{1}{2}\pm S_z)=(\frac{1}{2}\pm S_z)S_\mp=0\\ 
(\frac{1}{2}\pm S_z)S_\pm&=&S_\pm(\frac{1}{2}\mp S_z)=S_\pm\,.
\end{eqnarray}
Let us start by calculating the traces of products of lower--index matrices.
\begin{enumerate}
\item Consider a product of matrices that contains only factors $K_l$,
$\prod_{l=j}^kK_l$, where $j\ge k$ are two arbitrary positive integers. With
the above rules we easily find 
$\prod_{l=j}^kK_l=K_j\prod_{l=j}^{k+1}\alpha_l$ and
therefore, $\tr\prod_{l=N}^1K_l=\prod_{l=N}^{1}\alpha_l$.
\item A product containing only factors $L_l$ leads to
$\prod_{l=j}^kL_l=L_k\prod_{l=j}^{k+1}\beta_l$
and thus to $\tr\prod_{l=N}^1L_l=\prod_{l=N}^{1}\beta_l$.
\item A product $L_kK_j$ gives 
$L_kK_j=(\alpha_{j+1}\beta_k-\gamma_k^2)(\frac{1}{2}+S_z)$.
With this result we can calculate a product which contains $L$'s and $K$'s,
\begin{equation} \label{LK}
L_k\ldots L_jK_{j-1}\ldots K_i=(\prod_{l=k}^{j+1}\beta_l)(\alpha_j\beta_j-\gamma_j^2)(\prod_{l=j-1}^{i+1}\alpha_l)(\frac{1}{2}+S_z)\,,
\end{equation}
where $k\ge j\ge i$. Since $(\frac{1}{2}\pm S_z)^n=(\frac{1}{2}\pm S_z)$
for all integer $n$, we can easily generalize the above equation to products
containing several $L$ and $K$ blocks. We thus obtain the
following rule for the trace of an 
arbitrary product starting with $L_N$ and ending with $K_1$:
\begin{itemize}
\item Replace all $L_l$ by $\beta_l$ with the exception of the last
$L_l$ in each  $L$ block.
\item Replace all $K_l$ by $\alpha_l$ with the exception of the last
$K_l$ in each $K$ block.
\item Each interface between an $L$ block and a  $K$ block (in this
order, i.e.~a
term $L_jK_{j-1}$!) gives rise to a factor $\alpha_j\beta_j-\gamma_j^2$.
\end{itemize}
We will call these three rules the {\em standard rules}.
\item The next case to be considered is a product starting with $L_N$ and
ending 
with $L_1$, with an arbitrary number of $L_l$ and $K_l$ in between. We can
apply (\ref{LK}) to all blocks with the exception of the last $L$ block, to
which we apply directly the rule for a pure $L$ block. Besides the previous
replacement of 
$K_l$ by $\alpha_l$ and $L_l$ by $\beta_l$, one obtains a factor
$(\frac{1}{2}+S_z)L_1=L_1$. Since $\tr L_1=\beta_1$ the standard rules have
to be modified only in as 
much as $\tr L_N\ldots LK\ldots K\,\, \ldots L_1$ gives rise to an additional factor
$\beta_1$.  
\item Consider now a product starting with $K_N$ and ending with $L_1$. Due
to the cyclic property of the trace we can bring the last $L$ block to the
left (e.g. $\tr K_N\ldots K_jL_{j-1}\ldots L_1=\tr L_{j-1}\ldots L_1K_N\ldots
K_j$), thus creating an interface $L_1K_N$; and the product of matrices now
starts with an $L$ and  
terminates with a $K$ matrix. Thus, the $L$ and $K$ blocks and their interfaces
can  be replaced 
according to the standard rules. Due to the cyclic
properties of $K_l$ and $L_l$, also the interface $L_1K_N$ 
is of the type $L_jK_{j-1}$ with $j=1$ and is
therefore to be replaced by $\alpha_1\beta_1-\gamma_1^2$. In summary, the
standard rules apply, but we get an additional factor
$\alpha_1\beta_1-\gamma_1^2$. 
\item The last situation is a product starting with $K_N$
and ending with $K_1$. Here it is useful to treat the first block
$K_{N}\ldots K_{j}$ separately according to the rule for a pure $K$ block
and the subsequent sequence $L_{j-1}\ldots K_1$
according to (\ref{LK}). One therefore encounters a factor $\tr
K_N(\frac{1}{2}+S_z)=\alpha_1$. So again the standard rules apply, but we
get an additional factor $\alpha_1$.
\end{enumerate}

Let us now consider the corresponding expressions for the matrices with
upper indices resulting from the monodromy matrices.
\begin{enumerate}
\item A product containing only factors $L^{(l)}$ gives an expression with
the same structure as a product of factors $L_l$, 
$\prod_{l=j}^kL^{(l)}=L^{(k)}\prod_{l=j}^{k+1}\beta_l$.
It follows immediately that $\tr \prod_{l=N}^1L^{(l)}=\prod_{l=N}^{1}\beta_l$.
\item A product consisting only of factors $K^{(l)}$ behaves differently
from its counterpart with 
lower indices, for the last and not the first $K^{(l)}$ survives,
$\prod_{l=j}^kK^{(l)}=K^{(k)}\prod_{l=j}^{k+1}\alpha_l$.
However, since, $\tr K^{(1)}=\alpha_1$ we still have $\tr
\prod_{l=N}^1K^{(l)}=\prod_{l=N}^{1}\alpha_l$.
\item A product of $K^{(j)}$ and $L^{(l)}$ is now simpler with
$K^{(j)}$ on the left,
$K^{(j)}L^{(l)}=(\alpha_l\beta_l-\gamma_l^2)(\frac{1}{2}+S_z)-\beta_lS_\pm$.
We can combine the rules for the pure $K$ blocks, the pure $L$
blocks and
the $KL$--interface to derive the product 
\begin{equation} \label{KKLL}
K^{(k)}\ldots K^{(j)}L^{(j-1)}\ldots
L^{(i)}=\left(\prod_{l=k}^{j+1}\alpha_l\right)\left(\prod_{l=j-1}^{i+1}\beta_l\right)\left[(\alpha_i\beta_i-\gamma_i^2)(\frac{1}{2}+S_z)-\beta_iS_+\right]\,.
\end{equation}
If we have several such $KL$ blocks, each of them gives rise to a factor
corresponding to the one in angular brackets on the rhs of the
expression. But note that
$\left[(\alpha_i\beta_i-\gamma_i^2)(\frac{1}{2}+S_z)-\beta_iS_+\right]\left[(\alpha_l\beta_l-\gamma_l^2)(\frac{1}{2}+S_z)-\beta_lS_+\right]=(\alpha_i\beta_i-\gamma_i^2)\left[(\alpha_l\beta_l-\gamma_l^2)(\frac{1}{2}+S_z)-\beta_lS_+\right]$,
i.e.~from all $KL$ blocks only the last angular bracket survives whereas all
others are replaced by factors $(\alpha_i\beta_i-\gamma_i^2)$ where the
index is 
always the last index in the $L$ blocks. We thus recover the rule
of the 
replacement of $LK$ interfaces by factors
$(\alpha_i\beta_i-\gamma_i^2)$. In an arbitrary product starting with
$K^{(N)}$ and ending in $L^{(1)}$ the last angular bracket leads to the
factor $(\alpha_1\beta_1-\gamma_1^2)$; and all $K$ blocks, $L$ blocks, and $LK$
interfaces have to be replaced by the standard rules! Thus, we get just the
same expression as for a product starting with $K_N$ and ending with $L_1$. 
\item In a product starting with $K^{(N)}$ and ending with $K^{(1)}$ it is
useful to treat the last $K$ block separately. If $l$ is the last index of
an $L$ in the product, one is lead to a factor
$\tr((\alpha_l\beta_l-\gamma_l^2)(\frac{1}{2}+S_z)-\beta_lS_+)K^{(1)}=\tr(\alpha_l\beta_l-\gamma_l^2)K^{(1)}=(\alpha_l\beta_l-\gamma_l^2)\alpha_1$.
Thus the standard rules apply, but we get an additional factor $\alpha_1$,
in agreement with the result for the corresponding lower--index expression.
\item In $\tr L^{(N)}\ldots L^{(j)}K^{(j-1)}\ldots KL\ldots L^{(1)}$ we
treat the first $L$ block separately. One easily sees then that again the
standard rules apply up to an additional factor $\beta_1$, as was the case
for the corresponding expression with lower indices.
\item The last product to be considered is one that starts with $L^{(N)}$
and ends with $K^{(1)}$. We use the cyclic properties of the trace to shift
the first $L$ block to the right. Then (\ref{KKLL}) applies, and we see that
the last $L^{(j)}$ term on the right gives the same factor
$(\alpha_j\beta_j-\gamma_j^2)$ as did the $L_jK_{j-1}$ interface in the
corresponding expression with lower indices. The standard rules apply
without restriction.
\end{enumerate}
So we have shown that an arbitrary ordered product of factors $K^{(l)}$ and
$L^{(l)}$ ($l$ decreasing from $N$ to $1$ from left to right) has the same
trace as the corresponding product in which all $K^{(l)}$ are replaced by
$K_l$ 
and all $L^{(l)}$ by $L_l$. This proves that $\tr
\prod_{l=N}^1M_l=\tr \prod_{l=N}^1M^{(l)}$. 

The last thing to consider is the sign of each saddle point contribution for
arbitrary $N$. We will show now that it is always possible to choose all
minors of $\Q{2N}$ real and positive. Therefore the sign of each saddle point
contribution in (\ref{trPN}) is positive.\\
Observe first of all that without regularization all minors $D_l$ of
$\Q{2N}$ with the exception of the determinant $\det\Q{2N}$ itself are
zero. This is obvious for $l=1\ldots N$, since there the corresponding
matrix is part of the upper left zero block of $\det\Q{2N}$. For $l=N+m> N$
note that $D_l$ contains a $N\times N$ upper left block which is zero,
and a $N\times m$ $(\mu,\eta)$ block in the upper right corner. Upon
expanding $D_l$ after the 
first row one encounters subdeterminants  with a $(N-1)\times N$ upper
left zero block and a $(N-1)\times (m-1)$ upper right $(\mu,\eta)$
block. Both blocks together have $N-1$ rows in each of which only the $m-1$
elements on the right can be different from zero. Therefore the $N-1$ first
rows are always linearly dependent, unless $m=N$, the case that amounts to
the full determinant. Thus, all minors $D_l$ with $1\le l \le 2N-1$ are
zero.
 
Suppose now that we add to $\Psi_N$ a small quadratic term that vanishes and
has a maximum at the
saddle point $(\mbox{\boldmath $\mu$},\mbox{\boldmath
$\eta$})=(\mbox{\boldmath $\mu$}_{sp},\mbox{\boldmath 0})$, i.e.~a function
$-\epsilon\sum_{i=1}^N((\mu_i-\mu_{sp_i})^2+\eta_i^2)$ with infinitesimal
$\epsilon>0$. If the original integral is convergent, the small addition
will not change the  value of the integral in the limit $\epsilon\to 0$, but
allows us to determine the phase of all minors $D_l$. 
The matrices ${\bf
D}_l$ corresponding to $D_l$ get all replaced by ${\bf D}_l'={\bf D}_l
+\epsilon {\bf 1}_l$, where $ {\bf 1}_l$ is the unit matrix in $l$
dimensions. For $1\le l \le N$ we
are immediately lead to $D_l=\epsilon^l>0$. For $N+1\le l \le 2N-1$ we expand $D_l'$ in powers
of  $\epsilon$ and get 
$D_l'=D_l+\epsilon\,\tr {\bf D}_l+{\cal O}(\epsilon^2)=\epsilon\,\tr {\bf
D}_l+{\cal O}(\epsilon^2)$. To determine the traces $\tr {\bf D}_l$ we need
the second derivatives in the $(\eta,\eta)$ block of $\Q{2N}$. They are
given by
\begin{eqnarray}
\partial_{\eta_k}\partial_{\eta_l}\Psi_N&=&\left(\partial_{\eta_k}^2R(\mu_k,\nub_{k+1};\eta_k)+4\frac{(\dnmS{k})^2}{\partial^2_{\nub_k}R(\mu_{k-1},\nub_{k};\eta_{k-1})}+4\frac{(\dnnS{k-1})^2}{\partial^2_{\nub_{k-1}}R(\mu_{k-2},\nub_{k-1};\eta_{k-2})}
\right)\delta_{k,l}\nonumber\\
&&+4\frac{(\dnnS{k-1})(\dnmS{k-1})}{\partial^2_{\nub_{k-1}}R(\mu_{k-2},\nub_{k-1};\eta_{k-2})}\delta_{k-1,l}+4\frac{(\dnmS{k})(\dnnS{k})}{\partial^2_{\nub_k}R(\mu_{k-1},\nub_{k};\eta_{k-1})}\delta_{k+1,l}\,.
\end{eqnarray}
Since $\partial_{\eta_k}^2R(\mu_k,\nub_{k+1};\eta_k)<0$ 
and $\partial_{\nub_k}^2R(\mu_{k-1},\nub_{k};\eta_{k-1})<0$ for all $k$ at
the saddle point, 
the  diagonal elements of the $(\eta,\eta)$ block of $\Q{2N}$ are 
all real and positive definite (remember that we pulled out a minus sign in the
definition of 
$\Q{2N}$ in terms of second derivatives). Therefore $\tr {\bf D}_l$ is real
and larger than zero for all $l$.
Since also $\det \Q{2N}$
is real and positive, the sign of each saddle--point contribution is now
determined to be positive (cf.(\ref{arg})).
The proof of our trace formula (\ref{trPN}) for the $N$th trace is herewith
complete.  It is worth mentioning that the comments made about the decoherence
built in in the formula (\ref{eq:trP1fin}) for $\tr P$, the limit $\tau\to
0$, and the 
generality of the formula  apply to (\ref{trPN}) for $\tr P^N$as well .

\section{Evaluation for an integrable dissipative kicked top}\label{secKic}
We now propose to evaluate explicitly $\tr P^N$ for our integrable
dissipative kicked top. 
To do so, we need the periodic points of the classical map $P_{cl}$ and their
stabilities. 
\subsection{Periodic points}
Let us start by determining the fixed points of $P_{cl}$.
It is most convenient to work with Cartesian
coordinates,
$x=J_x/J$, $y=J_y/J$, and $z=J_z/J=\mu$ (with $x^2+y^2+z^2=1$). The
coordinates   $x_N'$, $y_N'$
and $z_N'$ immediately after the $N$th rotation are related to those
immediately before the $N$th rotation ($x_N$, $y_N$ and $z_N$) by
\begin{eqnarray}
x_N'&=&x_N\cos\beta+z_N\sin\beta\\
y_N'&=&y_N\\
z_N'&=&-x_N\sin\beta+z_N\cos\beta\,.
\end{eqnarray}
The dissipation takes these intermediate coordinates to those immediately
 before the $(N+1)$st rotation,
\begin{eqnarray}
x_{N+1}&=&\frac{x_N'}{\cosh\tau-z_N'\sinh\tau }\\
y_{N+1}&=&\frac{y_N'}{\cosh\tau-z_N'\sinh\tau }\label{yn}\\
z_{N+1}&=&\frac{-\sinh\tau+z_N'\cosh\tau }{\cosh\tau-z_N'\sinh\tau}\,.
\end{eqnarray}
For $\tau=0$, the dissipative part becomes the identical map and the fixed
points are those of the rotation, thus ${\bf x}_{N+1}={\bf x}_{N}=(0,\pm
1,0)$ for all $\beta$. \\
For $\tau\ne 0$, the equation for the $y$ component demands $y_N=0$ or
$\cosh\tau-  z_N'\sinh\tau=1$. Let us consider the latter equation 
first. It leads to the fixed points
\begin{equation} \label{fps}
x_N=-\cot\frac{\beta}{2}\tanh\frac{\tau}{2}\mbox{,}\quad y_N=\pm\sqrt{\frac{1-\cos\beta-2\tanh^2\frac{\tau}{2}}{1-\cos\beta}}\mbox{,}
\quad z_N=-\tanh\frac{\tau}{2}\,.
\end{equation}
They can exist with real $y_N$ only  if
$\sinh\frac{\tau}{2}\le|\tan\frac{\beta}{2}|$. If we compare (\ref{fps})
with ${\bf 
x}_{t+1}={\bf x}_{t}=(0,\pm 1,0)$ at $\tau=0$, we see that the pair of
fixed points shifts downwards in $z$
and towards the $y=0$ plane as $\tau$ increases. When $\tau=\tau_c$ given by
$\sinh\frac{\tau_c}{2}=|\tan\frac{\beta}{2}|$, the two fixed points
coincide in the plane $y=0$; a bifurcation takes place and
one has to consider $y_N=0$. \\
This case is treated most easily in polar coordinates, since $y_N=0$ means
$\phi=0$ and ${\bf x}$ depends only on the polar angle $\Theta$. The rotation becomes trivial, $\Theta'=\Theta_N+\beta$,
where  $\Theta$ should be allowed to run through $-\pi...\pi$ in order to
avoid $\phi=\pi$. The overdamped-pendulum trajectory (\ref{eq:sldmp})  implies
\begin{equation} \label{eq:tantheta}
\tan\frac{\Theta_{N+1}}{2}=\re^\tau\tan\frac{\Theta'}{2}=\re^\tau\tan\frac{\Theta_N+\beta}{2}
\end{equation}
throughout the interval
$\Theta=-\pi...\pi$. With the substitution $u_N=\tan\frac{\Theta_N}{2}$ the
periodic-point condition can be written as a 
quadratic equation, 
\begin{equation} \label{eq:ppc}
\tan\frac{\beta}{2}\,u_N^2+u_N(1-\re^\tau)+\tan\frac{\beta}{2}\re^\tau=0\,.
\end{equation} 
The two solutions 
\begin{equation} \label{eq:solppc}
u_{N_{1,2}}=\frac{\re^{\tau/2}}{\tan(\beta/2)}\left(-\sinh\frac{\tau}{2}\pm\sqrt{\sinh^2\frac{\tau}{2}-\tan^2\frac{\beta}{2}}\right)
\end{equation}
are real iff
$\sinh\frac{\tau}{2}\ge|\tan\frac{\beta}{2}|$,
i.e.$\tau>\tau_c$. Transforming $u_N$ back 
to $\mu$, we get
\begin{equation} \label{eq:solppc2}
\mu=\pm\cos\frac{\beta}{2}\sqrt{1-(\coth\frac{\tau}{2}\,\sin\frac{\beta}{2})^2}-\coth\frac{\tau}{2}\,\sin^2\frac{\beta}{2}\,.
\end{equation}
For $\tau\to\infty$, the lower one of the two fixed points moves to the
south pole $\mu=-1$ for all values of $\beta$. The upper one moves
to $\mu=-\cos\beta$, such that the intermediate point immediately after the
rotation is at the north pole, $\mu=1$. The final momentum $\nu$ is obtained
from (\ref{eq:solppc2}) by reversing 
the sign in front of the last $\coth\frac{\tau}{2}$. \\
The boundary between classically allowed
and forbidden rotation is  given by $\Theta_N+\Theta'=\beta$ (see section \ref{secuni}), or 
\begin{equation} \label{eq:bound}
\sin^2\beta-\mu^2-\nu^2+2\mu\nu\cos\beta=0\,.
\end{equation} 
 One easily checks
that the two fixed points given by (\ref{eq:solppc2}) have the remarkable
property that they lie exactly on this boundary, for all values of
$\tau>\tau_c$ and $\beta$, as 
long as the fixed points exist. Unfortunately, in this case the
semiclassical expression (\ref{eq:dmn}) for the unitary propagator ceases to
be valid. To avoid this problem we will  restrict ourselves to
$\tau<\tau_c$.\\
We show in appendix \ref{appFP} that $P_{cl}^N$ has the same and only the
same fixed points as $P_{cl}$. Therefore the restriction $\tau<\tau_c$
is sufficient as well for all higher traces. We also remark that $N\beta$
should not be an integer multiple of $2\pi$, otherwise already the
semiclassical approximation of the unitary part breaks down \cite{PBraun96}.
\\ 

In FIG.\ref{figphase} we show phase--space portraits. The northern and
southern hemisphere of the spherical phase space 
spanned by the variables $\mu$ and $\phi$ were
projected on the equatorial plane $\mu=0$. The two symmetry related fixed
points are clearly visible 
for $\tau<\tau_c$ as {\em elliptic fixed points}, whereas for $\tau>\tau_c$
they become a {\em pair of a point attractor and a point repeller}.

\subsection{Stability matrices}
The only non--trivial element of the monodromy matrix $M_d$ for the dissipative
motion is given by
$\frac{\partial\mu(\nu,\phi)}{\partial\nu}=\frac{d\mu_d}{d\nu}=
\frac{1-\mu^2}{1-\nu^2}$, 
whereas $\frac{\partial\phi_d(\nu,\phi)}{\partial\phi}=1$ and the two other
elements are zero. 
For the fixed point at $\tau<\tau_c$, $\nu=-\mu$, even this
non--trivial element equals one, and thus $M_d={\bf 1}$. In the vicinity of its
two fixed points our dissipative map is actually area--preserving. \\
The elements of $M_r$ can be found from differentiating the equations of
motion for the rotation written in $\mu,\phi$ and $\mu'=\nu,\phi'$,
\begin{eqnarray}
\sqrt{1-\nu^2}\sin\phi'&=&\sqrt{1-\mu^2}\sin\phi\\
\nu&=&\mu\cos\beta-\sqrt{1-\mu^2}\cos\phi\sin\beta\,.
\end{eqnarray}
We then obtain at $\nu=-\mu$
\begin{eqnarray}
\frac{\partial \phi'}{\partial
\phi}\left.\right|_\mu&=&1-\frac{\sin^2\beta}{1+\cos\beta}\sin^2\phi\\
\frac{\partial
\mu'}{\partial\mu}\left.\right|\phi&=&\cos\beta+\frac{\sin^2\beta}{1+\cos\beta}\cos^2\phi\,, 
\end{eqnarray}
and thus at the fixed points (\ref{fps}), 
\begin{equation} \label{eq:trdiff}
\tr M=\tr
M_1=\frac{2}{1-\tanh^2\frac{\tau}{2}}(\tanh^2\frac{\tau}{2}+\cos\beta)\,. 
\end{equation}
If we insert this and $\tr M_d=2$ in  (\ref{eq:trP1fin}), we obtain  the final result for $\tau<\tau_c$,
\begin{equation} \label{eq:trPfin}
\tr P=\frac{1-\tanh^2\frac{\tau}{2}}{1-\cos\beta-2\tanh^2\frac{\tau}{2}}\,.
\end{equation}
A factor 2 has canceled since two orbits contribute with equal weight. 
Clearly, $\tr P$ diverges at the bifurcation.\\
Since $M_1$ is unitary for $\frac{\partial\mu_d}{\partial \nu}=1$, its
eigenvalues can be written as 
$\lambda_{1/2}=\re^{\pm \ri\delta}$. We have therefore $\tr
M_1=2\cos\delta$ and  $\tr M_1^N=\cos(N\delta)=\cos(N\arccos \frac{\tr
M_1^N}{2})$, or 
\begin{equation} \label{trPNfin}
\tr P^N=\frac{1}{1-\cos(N\arccos(\frac{\tanh^2\frac{\tau}{2}+\cos\beta}{1-\tanh^2\frac{\tau}{2}}))}\,.
\end{equation}

\subsection{Comparison with numerics}
In FIG.\ref{figtrP} we have plotted  the first trace as semiclassically
approximated by (\ref{eq:trPfin}) together with 
the numerically obtained quantum counterpart for different values
of $j$. The agreement is  good for $\tau\gtrsim 1/J$ and
$\tau$ not too close to the critical $\tau_c$, where the bifurcation takes
place and the semiclassics diverges; the agreement improves with increasing
$j$. The rather 
erratic behavior of $\tr P$ as a function of $J$ when $\tau\lesssim 1/J$ is
a signature of the importance of quantum effects in the dissipative part. \\

In FIG.\ref{figtrPN} we compare the
semiclassical and quantum
mechanical ($j=20$) result for $\tr P^N$ as a function of
$N$. The agreement is rather good for  the
first about 10 traces. Both results show oscillations, and the semiclassical
result reproduces perfectly their period and phase  over
the entire $N$ range. However, the exact traces
decay exponentially as $\tr P^N\stackrel{N\to \infty}{\longrightarrow}1$,
whereas the  
semiclassical approximation (\protect\ref{trPNfin}) shows undamped
oscillations. Note that all  eigenvalues $\lambda_i$ of $P$ are bound by
$|\lambda_i|\le 1$ due to stability reasons. In general one has even
$|\lambda_i|< 1$ with the exception of one eigenvalue $\lambda_1=1$ which
corresponds to a stationary state. Since $\tr
P^N=\sum_{i=1}^{(2j+1)^2}\lambda_i^N$ the exact traces must indeed decay
exponentially towards $1$.

We have studied the decay of the exact traces   numerically in more
detail. It is an 
exponential decay of the form 
$\tr P^N\simeq 1+const.\re^{\frac{-f(\tau,\beta)N}{J}}$, where $f(\tau,\beta)$
is some function of the parameters. Thus, the effect of
the dissipation is very small and  vanishes for fixed finite $N$ in the
classical limit. 
Its onset should be visible one order beyond our present semiclassical
approximation. 
Note that the absence of classical periodic orbits with phase--space
contraction alone is not enough to expect a spectrum corresponding to a unitary
map (i.e. eigenvalues on the unit circle). This is due to the fact that
beyond the classical information entering via the periodic points 
our trace formula always contains a dissipative effect on the
quantum mechanical level, namely the complete destruction of coherence
manifest in the single sum over periodic points and the cancellation of the
actions from the unitary part.  

In order to arrive at the precision at which the exponential decay should be
seen, it would be necessary
to obtain at least the purely dissipative propagator $D$ to accuracy ${\cal
O}(1/J)$. So far our semiclassical 
calculations 
of $\tr P$ have been restricted by the requirement $1/J\ll \tau\ll J$ which
marked the limit of validity of the semiclassical approximation of $D$.
For the calculation of $\tr P^N$ one should have correspondingly $N/J\ll
N\tau\ll J$. 

In order 
to reconstruct the whole spectrum one would need traces from $N=1$ to
$N=(2J)^2$, such that $\tau$ could only be less than $1/J$ to fulfill the
last condition. But this conflicts with $\tau\gg 1/J$. We conclude
therefore that without pushing semiclassics one order further, it will never
be possible to calculate a whole spectrum. 
However, this might be a too
ambitious goal anyway: Even for moderate dissipation ($\tau\simeq 2$ for
example), most eigenvalues are exponentially small and can not be
reconstructed numerically even if one knew the exact traces! We tested this
 by calculating traces from exact quantum mechanical eigenvalues. From the
traces one can get the coefficients of the characteristic polynomial via  
Newton's formulae and therefore in principle back the eigenvalues. But
the polynomial becomes exponentially 
small for its argument close to zero, which makes it virtually impossible to
accurately determine its roots. For large values of $J$ one faces moreover
the problem that the higher traces ($N\sim J^2$) are all equal to unity up to
tiny corrections of order ${\cal O}(\exp(-J))$ and therefore 
carry no more  information. 
\\
On the other hand, it is quite likely that in typical experimental
situations only the slowest decaying eigenmodes and corresponding
eigenvalues can be measured. Future investigations will have to show, in
how far these can be obtained from a limited number of traces.

\section{Conclusion}\label{sectconcl}
We have established a semiclassical periodic--orbit theory for a dissipative
quantum map
with an area--nonpreserving classical limit. 
We have shown that traces of arbitrary integer powers of the 
propagator of the density matrix
can be written as a sum over classical periodic points.
Our results
generalize the well known formula due to  Tabor for area--preserving maps and
are in the spirit of Gutzwiller's trace formula.  Both the exponent and
preexponential factor in Tabor's formula 
are modified.  A ``diagonal approximation'' \cite{Hannay84,Berry85} which 
amounts to neglecting   interferences between different orbits in suitably
averaged quantities arises
automatically for our dissipative case: Interferences between different
orbits are destroyed by the 
dissipation. Therefore the exponent in the trace formula is purely real. Only
the action from the
dissipative part of the map remains. Even though only classical information
is needed to evaluate our trace formula, the quantum mechanical effect of
the destruction of coherence  by dissipation is automatically built in.
In the preexponent  the total monodromy
matrix, but also the monodromy matrix of the dissipative parts alone
enter. The Maslov index is always zero.\\

Our theory allows to quantify the  effects of the dissipation on the
spectrum of the propagator even when dissipation is so strong that it not
only leads to dephasing but also to a modification of periodic orbits.
Our example of
a  dissipative integrable kicked top showed, however,
that even though the classical map may be  area--nonpreserving
globally, it might have periodic orbits close to which it {\em is}
area--preserving. If only such orbits exist, the only influence of the
dissipation 
on the spectrum is via the  destruction of interferences between different
periodic orbits.

{\it Acknowledgments: }
We would like to thank J.Weber for producing the phase--space
portraits. Discussions with M.Ku\'s at early stages of this work are gratefully
acknowledged. D.B. would like to thank P.B. for hospitality during his stay in
St.Petersburg. P.B. thanks the staff of
the  Department of Theoretical Physics of the Essen University for their
hospitality and help during his stay in Germany; he acknowledges the
financial support of the Russian Foundation  for Fundamental Research, grant
No. 96-02-17937.  
\begin{appendix} 
\section{Saddle--point method for a complex function of several variables}
\label{appSaddle}
Let $\bx$ represent $M$ real variables $x_1\ldots
x_M$, and let $F(\bx)$ and $G(\bx)$ be complex valued functions of
$\bx$ with $\Re F\le 0$ in the 
volume  $V$ in ${\cal R}^M$ over which we are going to integrate. Suppose
that $V$ contains a single non--degenerate stationary
point $\bx_0$ with $\partial_{x_i} F(\bx_0)=0$, $i=1\ldots M$. Let us denote  by $Q_{M\times
M}$ the matrix of the negative second derivatives, $(Q_{M\times
M})_{ik}=-\partial_{x_i}\partial_{x_k}F(\bx)$ taken at $\bx=\bx_0$. The
condition of 
non--degeneracy means $\det \Q{M}\ne 0$. 
We then have for $J\to\infty$  \cite{Prudkovsky74}
\begin{equation} \label{eq:SPA}
\int_V
d^M\bx {\rm e}^{JF(\bx)}G(\bx)=G(\bx_0)\sqrt{\frac{(2\pi)^M}{J^M|\det
Q_{M\times 
M}|}}{\rm e}^{JF(\bx_0)-\frac{{\rm i}}{2}{\rm Ind} Q_{M\times M} }(1+{\cal
O}(1/J))\,. 
\end{equation}  
Here ${\rm Ind} Q_{M\times M}$ is the index of the complex quadratic form
$\chi$ of $M$ real variables, 
\begin{equation} \label{chi}
\chi=\sum_{i,j=1}^M(Q_{M\times M})_{ij}x_ix_j\,.
\end{equation}
The index is defined via the minors $D_k=\det||(Q_{M\times M})_{ij}||$, $1\le
i,j\le k$ of  $Q_{M\times M}$ as
\begin{eqnarray} \label{arg}
{\rm Ind}Q_{M\times M}&=&\sum_{k=1}^M\arg\rho_k\mbox{, }-\pi<
\arg\rho_k\le \pi\,,\\
\rho_1&=&D_1=(Q_{M\times M})_{11}\mbox{, }\rho_k=\frac{D_k}{D_{k-1}}, \quad
k=2\ldots M\,.
\end{eqnarray}
The restriction $-\pi\le
\arg\rho_k\le \pi$ on the phases fixes uniquely the overall phase of the
saddle point contribution. Without this restriction ${\rm Ind}Q_{M\times M}$
would only be defined up to multiples of $2\pi$, and this would lead to an
overall phase ambiguity corresponding to the choice of sign for the square
root in (\ref{eq:SPA}).

In the case of a $D_k=0$ we add  a term $-\epsilon (x_i-x_{0_i})^2$ to the
function $F(\bx)$. Such an addition does not change the  convergence
of the integral (in the limit of $J\to\infty$), nor its value for $\epsilon\to
0$. However, such a small term may bring the $D_k$ away from zero and
therefore allow us to determine its  phase.

Formula (\ref{eq:SPA}) has the structure familiar from the SPA of a complex
function of a single variable. The term $-\frac{{\rm i}}{2}{\rm Ind}
Q_{M\times M}$ 
leads to 
 a phase analogous to Maslov's $\pm\pi/4$, but that phase can now take on
any value between $-\frac{\pi}{2}...\frac{\pi}{2}$.  

\section{The determinant of a tridiagonal,
periodically continued matrix}\label{appdet}
The $N\times N$ matrix whose determinant we want to calculate has the structure
\begin{equation} \label{A}
A=\left(
\begin{array}{ccccccc}
a_{11}&a_{12}&0&0&\ldots&0&a_{1N}\\
a_{21}&a_{22}&a_{23}&0&\ldots&0&0\\
0&\ddots&\ddots&\ddots&\multicolumn{2}{c}{}&0\\
\vdots&&&&&&\vdots\\
a_{N1}&0&\multicolumn{3}{c}{\dotfill}&a_{N,N-1}&a_{NN}
\end{array}
\right)
\end{equation}
We first consider the simpler problem where $a_{1N}=a_{N1}=0$; the
corresponding matrix will be called $A_0$. The determinant of $A_0$ can be
expanded along the diagonal by using the block decomposition rule for
determinants \cite{Gantmacher86}. If $A$ is decomposed into smaller matrices
$W$ ($n\times n$), $X$ ($n\times m$), $Y$ ($m\times n$), and $Z$ ($m\times
m$, where $n+m=N$) and if the inverse of  $W$ exists, we have
\begin{equation} \label{detrel}
\det\left(\begin{array}{cc}
W&X\\
Y&Z
\end{array}
\right)=\det W\det(Z-YW^{-1}X)\,.
\end{equation} 
We apply the rule to $\det A_0$ by choosing $W=a_{11}$. Then $Z\equiv A^{(1)}$
is the matrix in which the first column and row are missing. In the matrix
$YW^{-1}X$ only one element is different from zero, namely
$(YW^{-1}X)_{ik}=\delta_{i,2}\delta_{k,2}\frac{a_{21}a_{12}}{a_{11}}$. Note
that we keep
the numbering of indices of the original matrix $A$. Thus, in the
determinant on the rhs of (\ref{detrel}), which is now only $(N-1)\times
(N-1)$,  the indices run from $2$ to $N$. The new matrix whose determinant
has now to be calculated differs from the
corresponding 
block in $A$ just in the upper left element $a_{22}$, and is 
tridiagonal again. The procedure may therefore be iterated. In each step the
dimension  of the matrix is decreased by one; the only matrix element
changed is always the one in the upper left corner. It gets replaced by
the original one in $A_0$ minus the product of the two secondary diagonal
elements flanking it divided by the upper left element of one step
earlier. One   obtains therefore a
product of renormalized diagonal elements,
\begin{equation} \label{proda0}
\det A_0=\prod_{j=1}^{N}c_j
\end{equation}
with $c_1=a_{11}$ and the iteration law 
\begin{equation} 
c_j=a_{jj}-\frac{a_{j,j-1}a_{j-1,j}}{c_{j-1}}\,.
\end{equation}
If we evaluate the product starting from the last term, we find that all
$c_j$'s with the exception of the one with the smallest index cancel, such
that  
\begin{equation} \label{prodan}
\prod_{j=n}^Nc_j=c_nf_n+g_n\,.
\end{equation}
The functions $f_n$ and $g_n$ depend only on the matrix elements, not
on the $c_j$'s. In the notation of (\ref{prodan}) $\det A_0$ is given by
$\det A_0=c_1f_1+g_1$, and the final condition for $f_n$ and $g_n$ reads
$f_N=1$, $g_N=0$. A recursion relation for $f_n$ and $g_n$ is easily
found by equating $\prod_{j=n-1}^Nc_j=c_{n-1}(c_n
f_n+g_n)=c_{n-1}(a_{nn}f_n+g_n)-a_{n,n-1}a_{n-1,n} f_n=c_{n-1}f_{n-1}+g_{n-1}$. We obtain a
linear relation which we write with the help of the matrix
\begin{equation} 
\bQ_n=\left(
\begin{array}{cc}
a_{nn}&1\\
-a_{n,n-1}a_{n-1,n}&0
\end{array}
\right)
\end{equation}
as
\begin{equation} \label{recurf}
{f_{n-1}\choose g_{n-1}}=
\bQ_n
{f_{n}\choose g_{n}}\,.
\end{equation}
The linear combination $c_1f_1+g_1$ at our interest therefore reads
\[
(c_1,1){f_1 \choose g_1}=(a_{11},1)\left(\prod_{j=2}^N\bQ_j\right) {f_{N}\choose g_{N}}\,.
\]
The product can be extended down to $j=1$ by noting that the vector
$(a_{11},1)$ is just the first row of the matrix in the product that would
carry the index 1. In the second row the product of elements
$-a_{1,0}a_{0,1}$ appears. Its value is at this stage completely
arbitrary. We  define it by periodically continuing the indices,
i.e.~$a_{1,0}\equiv a_{1,N}$ and $a_{0,1}\equiv a_{N,1}$. We then arrive at
the result for the determinant of the tridiagonal matrix $A_0$,
\begin{equation} \label{deta0fin}
\det A_0=(1,0)\left(\prod_{j=1}^N\bQ_j\right){1\choose 0}\,.
\end{equation}

Consider now the case where $a_{1,N}a_{N,1}\ne 0$. The problem of
calculating $\det A$ can be reduced to the problem of determinants of 
purely tri--diagonal matrices by expanding $\det A$ after the first row and the
resulting submatrices after the first column. One eliminates in this way the
rows and columns with the disturbing additional corner
elements. A matrix $A^{(1,N)}$ appears, which is defined as 
the matrix $A$ from which both the first and $N$th rows and columns have
been taken out. The 
reader will easily verify that 
\begin{equation} \label{deta1}
\det A=\det A_0-a_{1N}a_{N1}\det A^{(1,N)}+(-1)^{N+1}\left(\prod_{j=1}^Na_{j,j-1}+\prod_{j=1}^Na_{j-1,j}\right)\,.
\end{equation}
For $a_{1N}a_{N1}=0$ the expression reduces to $\det A_0$ as it should. It will
be useful to rewrite the sum of the two products as trace of a diagonal
$2\times 2$ matrix which is the product of matrices $\bR_j\equiv {\rm
diag}(a_{j,j-1},a_{j-1,j})$. 
Let
us apply now our result for tri--diagonal determinants (\ref{deta0fin}). We
obtain 
\begin{equation} \label{|A|almost}
\det A=(1,0)\left[\left(\prod_{j=1}^N\bQ_j\right)-a_{1,N}a_{N,1}\left(\prod_{j=2}^{N-1}\bQ_j\right)
\right]{1\choose 0}+(-1)^{N+1}\tr\prod_{j=1}^N\bR_j\,.
\end{equation}
Now observe that 
\begin{eqnarray}
(0,1)\bQ_1&=&-a_{1,N}a_{N,1}(1,0)\mbox{ and}\\
\bQ_N{0\choose 1}&=&{1\choose 0}\,.
\end{eqnarray}
Thus, while the first term in (\ref{|A|almost}) gives the upper left matrix
element of the product of matrices, the second one gives the lower right. We
therefore have
\begin{equation} 
\det A=\tr\prod_{j=1}^NQ_j+(-1)^{N+1}\tr\prod_{j=1}^N\bR_j\,.
\end{equation}
The order of the products is such that they start with the matrices with
index 1 on the left.  
Given the ordering of the monodromy matrices that appears in the main text
it is useful to inverse the order. This can easily be  done by taking the
traces of 
the transposed product, which leads to the final result
\begin{equation} \label{finaldet}
\det A=\tr\prod_{j=N}^1\left(
\begin{array}{cc}
a_{jj}&-a_{j,j-1}a_{j-1,j}\\
1&0
\end{array}
\right)+(-1)^{N+1}\tr\prod_{j=N}^1\left(
\begin{array}{cc}
a_{j,j-1}&0\\
0&a_{j-1,j}
\end{array}
\right)\,.
\end{equation}
Do not apply the formula for $N=2$, since we assumed {\em additional}
elements in the corners. For $N=2$ the corners collapse with the secondary
diagonal and are therefore over counted in (\ref{finaldet}). The lowest
dimension for which the formula works is $N=3$.

\section{Fixed points of $P_{cl}^N$}\label{appFP}
Let us first consider the case $y_n\ne 0$. Equating
$y_{N+1}$ with $y_1$ in (\ref{yn}), one is lead immediately to
$\prod_{i=1}^N(\cosh\tau-\sinh\tau z_i')=1$. Since the same denominator also
appears in $x_{N+1}$ and $z_{N+1}$ when expressed as functions of $x_1$ and
$z_1$, the remaining map $(x_1,z_1)\to (x_{N+1},z_{N+1})$ is in fact {\em
linear!} Foreseeing this we can 
build it up by iterating a linear map ${\bf A}$ which acts on the two
dimensional vector $\zeta_l\equiv(x_l,z_l)$ as 
\begin{equation} \label{lima}
\zeta_{l+1}={\bf A}\zeta_l+{\bf b}\,.
\end{equation}
The matrix ${\bf A}$ and the vector ${\bf b}$ are  explicitly given by 
\begin{equation} \label{mA}
{\bf A}=\left(\begin{array}{cc}
\cos\beta&\sin\beta\\
-\cosh\tau\sin\beta&\,\,\,\cosh\tau\cos\beta
\end{array}
\right)\mbox{, }{\bf b}={0 \choose -\sinh\tau}\,.
\end{equation}
The condition for the fixed points of $P_{cl}$ reads $({\bf
1}-\bA)\zeta_1=\bb$. The determinant of $({\bf 1}-\bA)$ is
$(1-\cos\beta)(1+\cosh\tau)$ and therefore non--zero as long as there is any
rotation at all. The equation is solved easily and gives back the fixed
point at $z_1=-\tanh\frac{\tau}{2}$, but additionally we now see directly
that it is 
unique in $x_1,z_1$, as the map is linear and the determinant non--zero. The
$y$ component is obtained from normalization, 
$y_1=\pm\sqrt{1-x_1^2-z_1^2}$. These are the two basic symmetry related
fixed points of $P_{cl}$. \\
The fixed points of $P_{cl}^2$ follow from $\zeta_1=\zeta_3=\bA
\zeta_2+\bb=\bA^2\zeta_1+({\bf 1}+\bA)\bb$, i.e. $\zeta_1=({\bf
1}-\bA^2)^{-1}({\bf 1}+\bA)\bb=({\bf 1}-\bA)^{-1}\bb$. One easily verifies
by complete induction 
that the formula for the $N$th iteration is
\begin{equation} \label{nth}
({\bf 1}-\bA^N)\zeta_1=(\sum_{i=1}^N\bA^i)\bb=\left(\frac{1-A^N}{1-A}\right)\bb
\end{equation} 
   and therefore again just $({\bf 1}-\bA)\zeta_1=\bb$, the equation for the
   fixed point of $P_{cl}$.\\

For $\phi=0$ the situation is similar: One easily verifies that 
$u_{N+2}=u_N$ in equation (\ref{eq:tantheta}) leads exactly to the same
equation (\ref{eq:ppc}) as did $u_{N+1}=u_N$, and the same is true for all
higher iterations. This completes our prove that $P_{cl}^N$ has the same and
only the same fixed points as $P_{cl}$.
\end{appendix}

\begin{figure}
\protect\caption{\label{figphase} Phase--space portraits of the classical
map of the dissipative  integrable kicked top ($\beta=1.2$). 
We projected the northern and the southern hemisphere ($\mu>0$ and $\mu<0$,
respectively) of the spherical classical phase space on the equatorial plane
$\mu=0$; (a) $0.6=\tau<\tau_c\simeq 1.28$, the two 
fixed points are elliptic 
fixed points; (b) $1.3=\tau>\tau_c\simeq 1.28$, the fixed points form   a point
attractor/repeller pair. The repeller in the northern hemisphere can be made
visible by iterating the inverse map.}  
\end{figure}

\begin{figure}
\protect\caption{\label{figtrP}} Comparison of semiclassical (dashed line)
and quantum 
mechanical result for $\tr P$ as a
function of the damping 
$\tau$ at  $\beta=2.0$. The semiclassical result (\ref{eq:trPfin}) diverges at
$\tau=\tau_c\simeq 2.45$, and the quantum mechanical trace follows this
divergence the 
further the higher the value of $J$ (circles $j=10$, squares $j=20$,
diamonds $j=40$). For $\tau\lesssim 1/J$ the problem is
purely quantum mechanical and  our semiclassical treatment seizes to be valid. 
\end{figure}

\begin{figure}
\protect\caption{\label{figtrPN}} Comparison of semiclassical and quantum
mechanical result for $\tr P^N$ ($\beta=0.5$, $\tau=0.2$) as a function of
$N$. FIG.(a) for $j=20$ shows that the agreement is fairly good for about the
first 10 traces and the quantum mechanical oscillations
are well reproduced over the entire $N$ range. However, the exact traces
decay, whereas the 
semiclassical approximation (\protect\ref{trPNfin}) only shows the
oscillations (same symbols as in FIG.1). In (b) the exponential decay of the
exact traces is studied numerically for
different values of $j$. It is of the form $\tr P^N\simeq
1+const.\exp(\frac{-N}{9.5J})$ for $\beta=0.5$ and $\tau=0.2$. 
\end{figure}

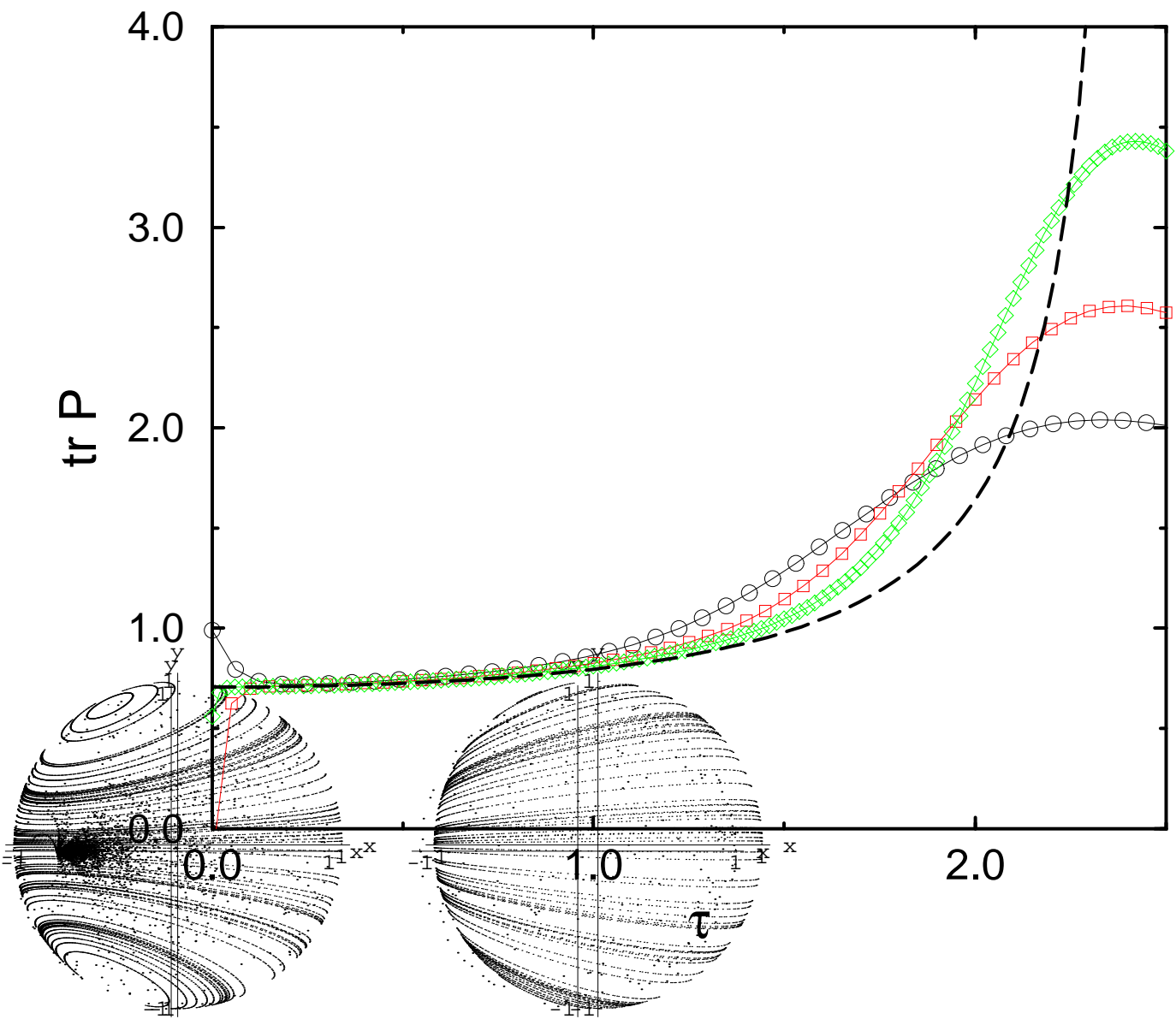
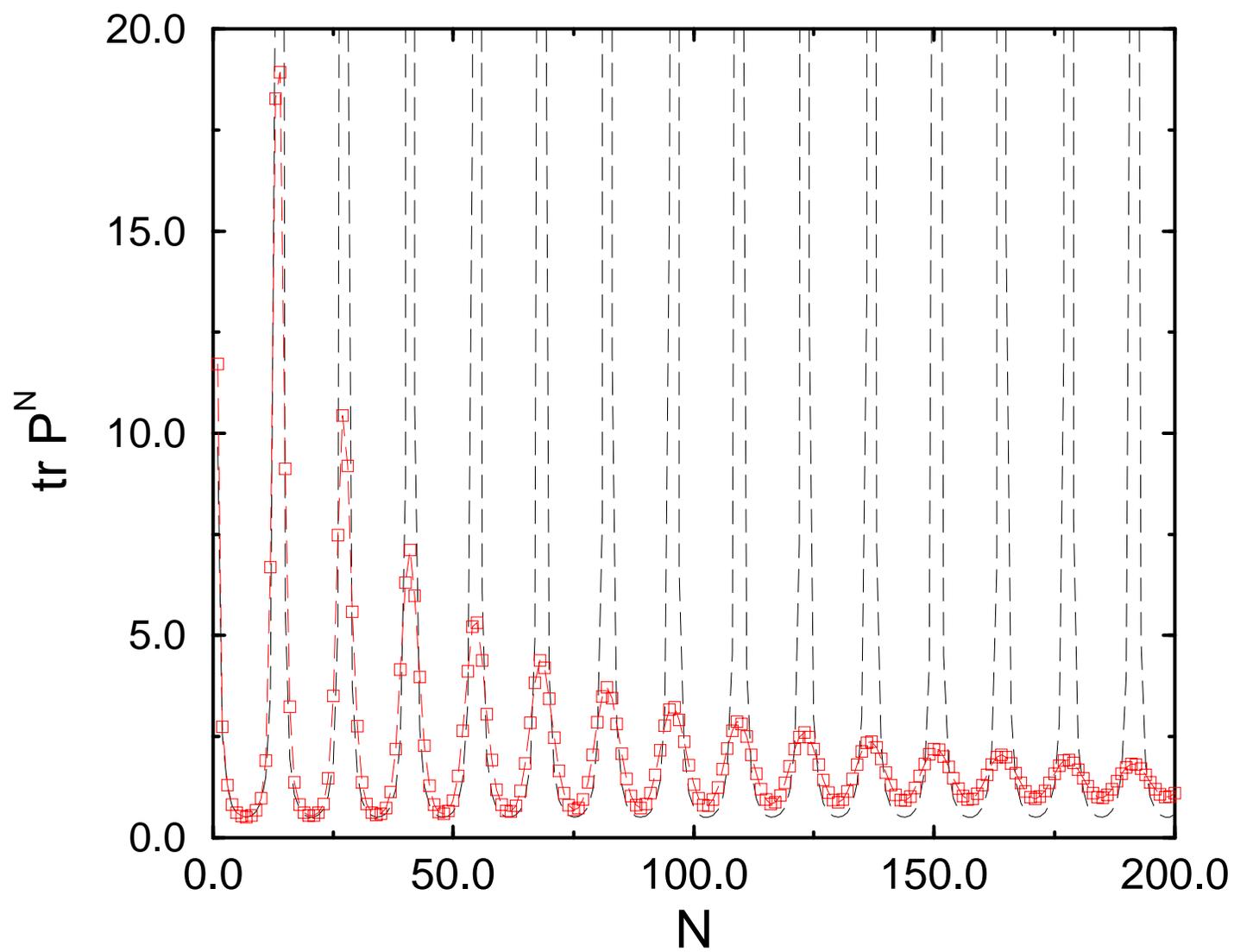
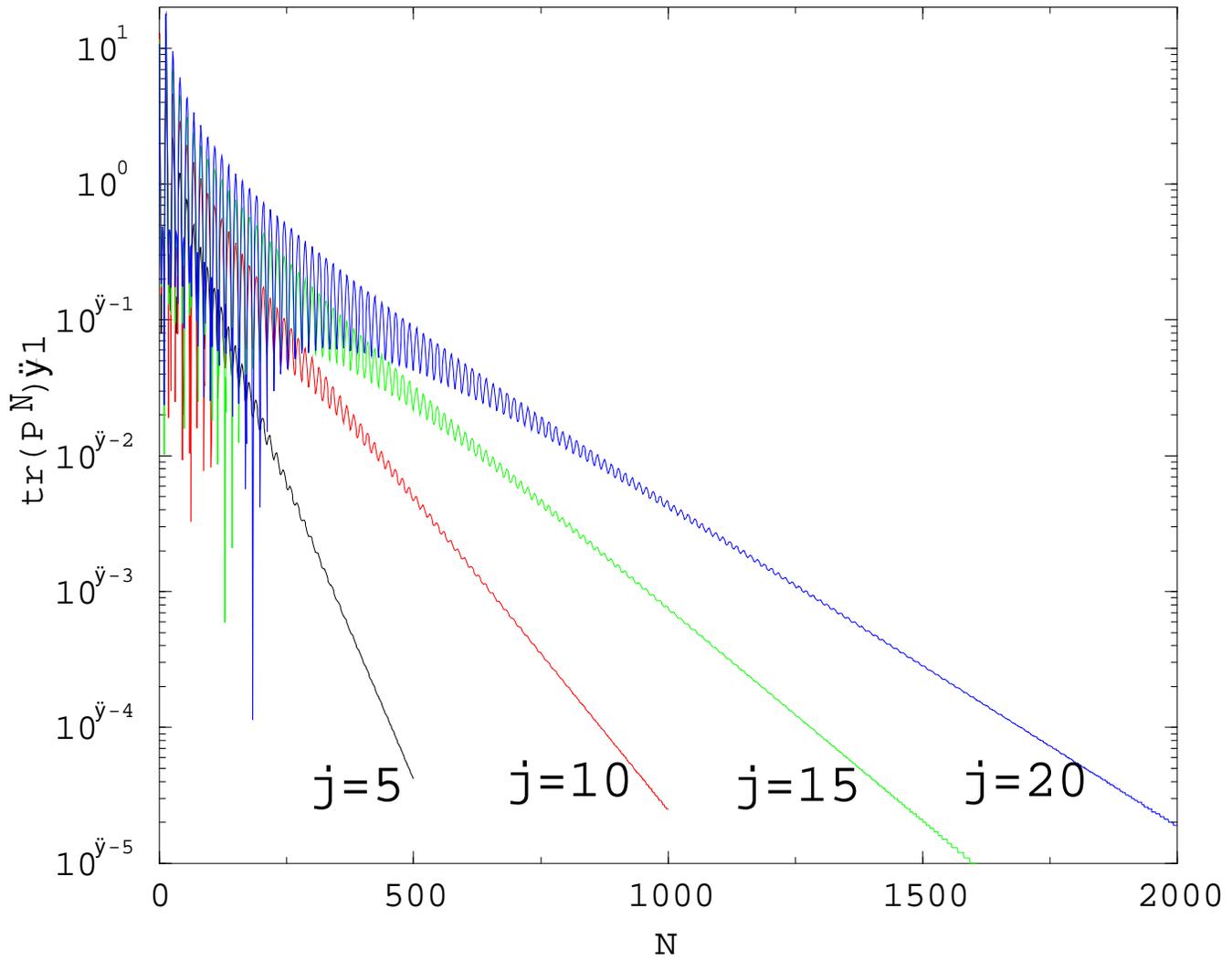
\end{document}